\DocumentMetadata{}
\documentclass[acmtog, authorversion]{acmart}

\AtBeginDocument{%
  \providecommand\BibTeX{{%
    \normalfont B\kern-0.5em{\scshape i\kern-0.25em b}\kern-0.8em\TeX}}}

\setcopyright{acmcopyright}
\acmJournal{TOG}
\acmYear{2025}
\acmVolume{44}
\acmNumber{4}
\acmArticle{0}
\acmMonth{8}
\acmPrice{xx}
\acmDOI{10.1145/3731181}

\usepackage{float} 
\usepackage{multirow}
\usepackage[ruled]{algorithm2e} 
\usepackage{wrapfig}
\usepackage{amsfonts}
\usepackage{amsmath}
\usepackage{algpseudocode} 
\usepackage{adjustbox}

\SetAlFnt{\small}
\SetAlCapFnt{\small}
\SetAlCapNameFnt{\small}
\SetAlCapHSkip{0pt}

\newtheorem{proposition}{Proposition}[section]

\usepackage{lipsum,graphicx,subcaption}
\newcommand{\revise}[1]{{\color{black} #1}}

\begin{document}

\begin{sloppypar}



\title{\revise{Designing 3D Anisotropic Frame Fields with Odeco Tensors}}



\author{Haikuan Zhu}
\email{hkzhu@wayne.edu}
\orcid{0000-0001-5740-0599}
\affiliation{
\institution{Wayne State University}
\country{USA}
  }

\author{Hongbo Li}
\email{hm9026@wayne.edu}
\orcid{0009-0005-9072-1489}
\affiliation{%
\institution{Wayne State University}
\country{USA}
}

\author{Hsueh-Ti Derek Liu}
\email{hsuehtil@gmail.com}
\orcid{0009-0001-1753-4485}
\affiliation{
\institution{Roblox \& University of British Columbia}
\country{Canada}
}

\author{Wenping Wang}
\email{wenping@tamu.edu}
\orcid{0000-0002-2284-3952}
\affiliation{
\institution{Texas A\&M University}
\country{USA}
  }

\author{Jing Hua}
\email{jinghua@wayne.edu}
\orcid{0000-0002-3981-2933}
\affiliation{%
\institution{Wayne State University}
\country{USA}
}

\author{Zichun Zhong}
\authornote{Corresponding author}
\email{zichunzhong@wayne.edu}
\orcid{0000-0001-6489-6502}
\affiliation{%
\department{Department of Computer Science}
\institution{Wayne State University}
\streetaddress{5057 Woodward Ave., Suite 14109.2}
\city{Detroit}
\state{MI}
\country{USA}
\postcode{48202}
}


\renewcommand{\shortauthors}{Haikuan Zhu et al.}

\newcommand{\norm}[1]{\left\lVert #1 \right\rVert}
\begin{abstract}
This paper introduces a method to synthesize a 3D tensor field within a constrained geometric domain represented as a tetrahedral mesh.
Whereas previous techniques optimize for \emph{isotropic} fields, we focus on \emph{anisotropic} tensor fields that are smooth and aligned with the domain boundary or user guidance. 
The key ingredient of our method is a novel computational design framework, built on top of the \emph{symmetric orthogonally decomposable} (odeco) tensor representation, to optimize the stretching ratios and orientations for each tensor in the domain. 
In contrast to past techniques designed only for \emph{isotropic} tensors, we demonstrate the efficacy of our approach in generating smooth volumetric tensor fields with high \emph{anisotropy} and shape conformity, especially for the domain with complex shapes.
We apply these anisotropic tensor fields to various applications, such as anisotropic meshing, structural mechanics, and fabrication.
\end{abstract}





\maketitle

\section{Introduction}
\begin{figure}
  \includegraphics[width=\linewidth]{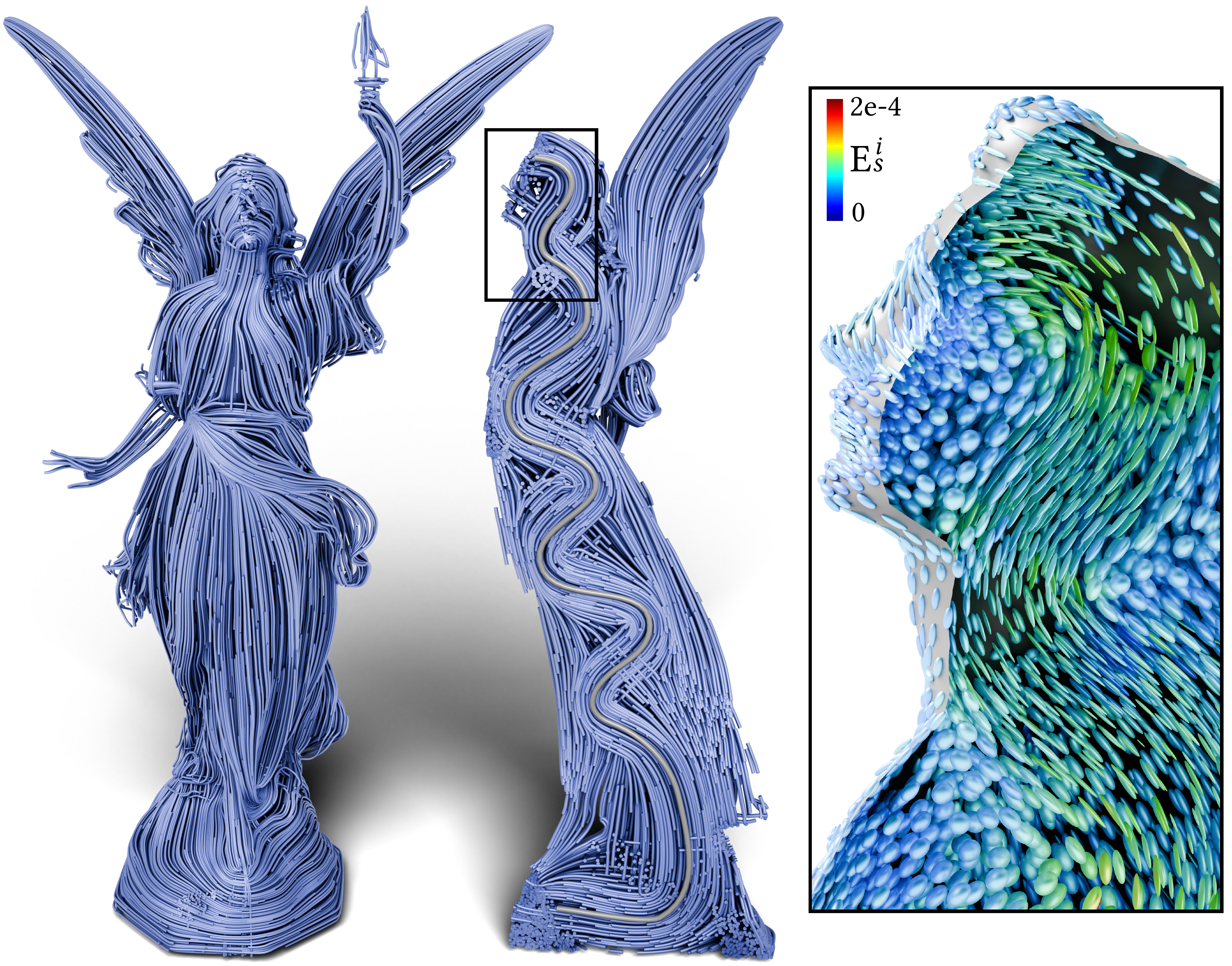}
  \caption{An optimal 3D anisotropic odeco tensor field (AOTF) is demonstrated through integral curves and clipped views. It automatically balances field smoothness, normal alignment, curvature alignment, feature preservation, and user-specified interior constraint (i.e., a gray curve shown in the middle image), leveraging the estimated surface curvature magnitudes as the user's input guidance. $E_{s}^i$ is the vertex-wise smoothness energy in this work.}
  \label{fig:teaser}
  
\end{figure}

Directional fields plays a central role in numerous graphics and engineering applications, ranging from texturing~\cite{ma2011discrete}, deformation~\cite{von2006vector}, to meshing~\cite{li2024nasm,zhong2018computing}. 
Although such a tensor field is relatively well-studied on \emph{surfaces} \cite{vaxman2016directional}, computing \emph{volumetric} tensor fields receives far less attention, despite its importance. 
Previous attempts, such as \cite{ray2016practical, corman2019symmetric, palmer2020algebraic}, propose techniques for computing volumetric tensor fields and demonstrate applications in e.g., hexahedral meshing. 
However, their formulations solely focus on \emph{isotropic} tensors, and fail to generate practical \emph{anisotropic} fields. This prevents them from being adapted to tasks such as volumetric computation on anisotropic problems \cite{loseille2014metric, yamakawa2000high} and the design of microstructures \cite{fabrication2019design, martinez2018polyhedral}.

In response, we propose an optimization framework for computing \emph{anisotropic} volumetric tensors. A key idea is to lean on the \emph{symmetric orthogonally decomposable} (odeco) representation~\cite{robeva2016orthogonal, boralevi2017orthogonal}. In contrast to the common symmetric matrix representation in \cite{palacios2016tensor}, the odeco representation provides an one-to-one mapping to the homogeneous polynomials. This property enables us to decompose a tensor to its orientation and the stretching components, which account for the direction and the anisotropy, respectively. 

We utilize the odeco parametrization to formulate a Dirichlet energy for optimizing anisotropic volumetric (tensor) frame fields. The separable stretching component enables users to control anisotropy for e.g., customizing anisotropic microstructure designs by simply assigning Young's moduli magnitudes only (see Fig.~\ref{fig:fab_hand}). 
We demonstrate that our optimization results in smooth \emph{anisotropic odeco tensor fields} (AOTF) that are naturally aligned with the domain boundary (see Fig.~\ref{fig:teaser}) both empirically and theoretically. 
This attribute offers benefits when introducing anisotropy in intricate geometries, and for other downstream applications, for instance, creating feature-preserving anisotropic meshes. \revise{Meanwhile, our anisotropic frame field design and optimization strategies in this work could provide a strong foundation to the challenging research in anisotropic hexahedral meshing as discussed in Section~\ref{se:future_work}.} Overall, our main contributions are:
\begin{itemize}
    \item We introduce a method for optimizing volumetric tensor fields with a novel focus on \emph{anisotropy}.
    \item Our optimization framework results in a smooth volumetric tensor field with high conformity to surface normals and curvatures, supported by our theoretical analysis.
    \item Our method incorporates user controls, demonstrating applications in anisotropic meshing, microstructure fabrication, and material design. 
\end{itemize}

\section{Related Work}
In this section, we review the related literature on cross and frame field, as well as 3D tensor field.

\subsection{Cross and Frame Field}
In geometric modeling and processing, there are lots of work in the design of direction fields in 2D and 3D domains, such as cross fields and their applications in quad meshing~\cite{vaxman2016directional}. Methods to compute smooth cross fields on the surface with alignment and singularity constraints are studied in~\cite{palacios2007rotational,ray2008n,knoppel2013globally,jakob2015instant,huang2016extrinsically,zhang2020featureframe,ma2024computing,dong2024neurcross}. Those works explore many approaches to achieve different geometric constraints, such as normal alignment, shape alignment, and feature alignment when smoothing the target direction fields. One of the most prominent representations is the trigonometric pair, ($sin4\theta,cos4\theta$) or the complex exponential $e^{i4\theta}$ equivalently. However, this representation does not encode scalings along with the directions. \cite{zhang2020featureframe} provides a useful theoretical analysis of normal-aligned SH-based octahedral frames on the surface, but is limited for orientations only. Inspired by that, we generalize their theoretical analysis to a more complicated case for volumetric tensor fields, where the stretching ratios are varied and distinct. There are few methods focusing on cross fields with the consideration of metrics. For instance, \cite{Panozzo:2014} introduces a non-orthogonal and non-unit-length generalization of cross fields and applies in high-quality adaptive quad meshing on surface. \cite{jiang2015frame} proposes frame field design constraints on alignment, size, and skewness on the mesh as well as along feature curves, offering more flexibility over previous approaches. \cite{simons2024anisotropy} considers a cross field with anisotropy and only works to regions in the plane with boundary.

However, it is usually not trivial to extend those ideas from surfaces to 3D volumes. The main geometric application for 3D frame fields is hexahedral and hex-dominant mesh generation~\cite{nieser2011cubecover,li2012all,lyon2016hexex,liu2018singularity,gao2017robust,sokolov2016hexahedral}. \cite{huang2011boundary} introduces a particularly convenient representation of a 3D octahedral frame as a rotation of the spherical function (SH). 

Inspired by the method of~\cite{huang2011boundary}, \cite{ray2016practical} produces a smooth 3D frame field, which varies smoothly inside the volume, and matches the normals of the volume boundary. \cite{solomon2017boundary} leverages the boundary element method (BEM) to achieve infinite resolution within the volume, enabling the continuous assignment of frames to points in the interior based solely on a triangle mesh discretization of the boundary. \cite{corman2019symmetric} proposes an alternative method based on Cartan's method of moving frames for computing a frame field with a prescribed singularity structure, utilizing a discretized connection on a frame bundle. \cite{fang2021metric} generates a smooth frame field for a Riemannian metric field that is flat away from singularities under a few intuitive constraints. But these representations of 3D cross and frame fields only consider the orientations. 

\subsection{3D Tensor Field}
Traditional cross and frame fields describe only orientations or directions within space. Research on those direction fields is unable naturally extend to the processing of tensor fields. This is because a 3D tensor contains not only directional information but also anisotropic scaling information (stretching ratios). 3D tensor fields have been widely used in different domains for decades, such as anisotropic meshing~\cite{yamakawa2000high,loseille2014metric,zhong2018computing}, microstructure design~\cite{fabrication2019design, martinez2018polyhedral}, deformation~\cite{zheng2002volume}, simulation~\cite{dick2009stress}, visualization~\cite{bi2019survey,hung2023global}, and mechanics~\cite{kratz2014tensor,hergl2021visualization}. However, computing and designing volumetric tensor fields has been little studied. Although \cite{palacios2016tensor} provides a tensor field design framework that specializes in the topology and singularities, the way of element-wise smoothing of the tensor matrix always leads to unsatisfying transitions. Besides, their normal alignment is achieved by tensor modification by force. Recently, benefiting from the symmetric orthogonally decomposable (odeco) tensor varieties introduced in~\cite{robeva2016orthogonal,boralevi2017orthogonal}, it allows for an exact one-to-one correspondence between symmetric tensors and homogeneous polynomials. This property enables us to decompose a tensor into its orientation and the stretching components. \revise{\cite{palmer2020algebraic} is the first to use anisotropic tensors to have energetically cheap singular curves}; however, their limited investigation of anisotropy results in a simple smoothed odeco tensor field with near uniform stretching ratios. Moreover, as observed in our experiments, the projection which is required in their method becomes unstable when flexible stretching ratios are introduced. Recently, \cite{couplet2024integrable} proposes integrable frame fields using odeco representation in both isotropic and anisotropic settings; however, they only focus on 2D planar frame field design. Extending their approach to 3D volumetric settings remains an open challenge. In this work, we address the challenging problem of designing 3D anisotropic odeco tensor fields in volume with the consideration of large stretching ratios in the company of additional boundary characteristics.

\section{Odeco Tensor Representation}
\label{se:algebraic}
Our odeco tensor spaces are defined on the volume of a tetrahedral mesh $\Omega = (\mathcal{V}, \mathcal{T})$ with vertices $\mathcal{V}$ and tetrahedra $\mathcal{T}$. The boundary (surface) domain is denoted by $\partial \Omega$. The odeco tensor field $f$ are discretized by set of orientations and stretching ratios defined on the vertices $\mathcal{V}$, denoted as $\boldsymbol{\Theta} = \boldsymbol{\theta}_1, \cdots, \boldsymbol{\theta}_{|\mathcal{V}|}$ and $\boldsymbol{\Lambda}= \boldsymbol{\lambda}_1, \cdots, \boldsymbol{\lambda}_{|\mathcal{V}|}$, respectively. For each odeco tensor $f(\boldsymbol{\theta}_i, \boldsymbol{\lambda}_i)$ at a vertex $i$ (abbreviated as $f_i$), $\boldsymbol{\theta}_i =(\theta_i^x, \theta_i^y, \theta_i^z) \in \mathbb{R}^3$ are \revise{Euler angles} to control the orientation and $\boldsymbol{\lambda}_i = (\lambda^x_i, \lambda^y_i, \lambda^z_i) \in \mathbb{R}^3$ are stretching ratios along each axis to control the anisotropy.

\subsection{Odeco Polynomial}
\revise{The symmetric orthogonally decomposable (odeco) tensor varieties introduced in \cite{robeva2016orthogonal} allow for an exact one-to-one correspondence between symmetric tensors $T= \sum_{i=1}^n \lambda_i v_i^{\bigotimes d} \in S^d(\mathbb{R}^n)$ of order $d$ and homogeneous polynomials $f$ of degree $d$ in $n$ variables $x=(x_1,...,x_n)$, where $v_i^{\bigotimes d}$ denotes the $d$-wise tensor power of the orthonormal vectors $v_i$. The weight of the odeco tensor $\lambda_i$ allows us to design independent stretching ratios.
$T$ is odeco if and only if the coefficients $u$ of the polynomial $f(x)=\sum u_{d_1,...,d_n} \Pi_{i=1}^n x_i^{d_i},\, \sum d_i=d$ satisfy $u^\intercal A_m u=0$ for a finite set of symmetric matrices $A_m$~\cite{palmer2020algebraic}. In this case, each odeco tensor $T$ is also referred to as an odeco polynomial, denoted as $f_i$.}

\subsection{Interior Odeco Tensor}
We decompose each odeco tensor $f_i$ into two components: one for the orientation $\boldsymbol{\theta}_i$ and one for the stretching $\boldsymbol{\lambda}_i$ as
\begin{equation}\label{eq:f_interior}
    f(\boldsymbol{\theta}_i,\boldsymbol{\lambda}_i)= \underbrace{e^{\theta_i^x \mathbf{L}_x} e^{\theta_i^y \mathbf{L}_y} e^{\theta_i^z \mathbf{L}_z}}_{\text{orientation}} \underbrace{\hat{f}(\boldsymbol{\lambda}_i)}_{\text{stretching}}, \quad\forall i\in \Omega,
\end{equation}
where the orientation component is a $\mathbb{R}^{15\times 15}$ rotation matrix based on the exponentiation of the Lie algebra elements \cite{palmer2020algebraic, zhang2020featureframe}, $\mathbf{L}_x,\mathbf{L}_y,\mathbf{L}_z \in \mathbb{R}^{15 \times 15}$ are the angular momentum operator, which can be found in Section 1.1 of Supplementary Material, and $\hat{f}$ is an odeco tensor on the canonical axes that account for the stretch. 
Intuitively, we define our odeco tensor space by rotating the canonical odeco tensor $\hat{f}(\boldsymbol{\lambda}_i)$ according to the rotation specified by $\boldsymbol{\theta}_i$.
This decomposition enables us to optimize the orientation $\boldsymbol{\theta}_i$ and anisotropy $\boldsymbol{\lambda}_i$ separately to obtain anisotropic tensor fields.  

\begin{figure}
    \centering
    \includegraphics[width=\linewidth]{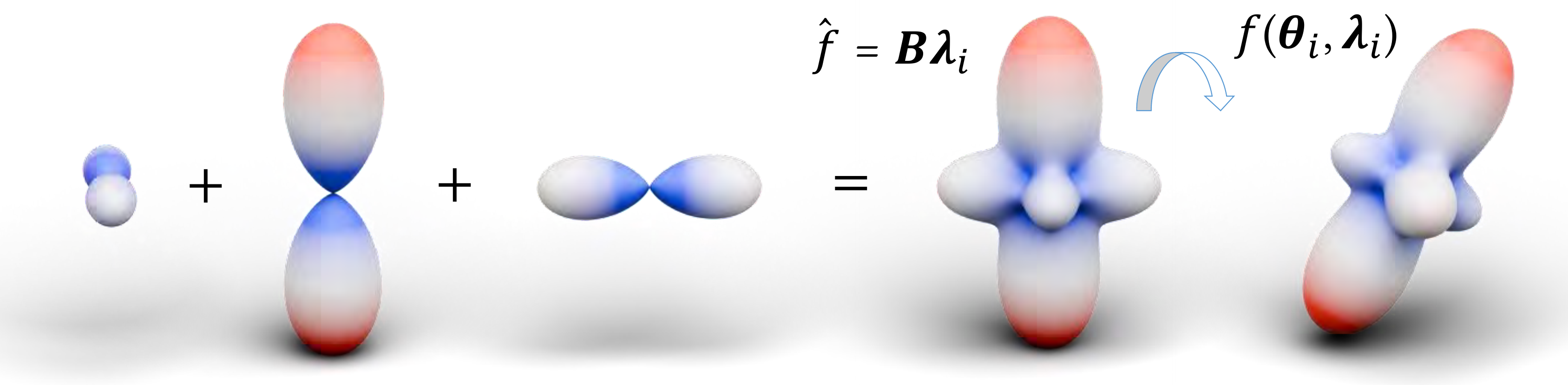}
    \caption{We represent each odeco tensor $f$ (right) as a rotation of the canonical odeco tensor $\hat{f}$ (middle). This canonical tensor $\hat{f}$ can be expressed as a linear combination of spherical harmonics bases and stretching ratios $\boldsymbol{\lambda}_i$ (left).}
    \label{fig:fz}
\end{figure}

In Eq.~(\ref{eq:f_interior}), the canonical $z$-axis-aligned odeco tensor $\hat{f}(\boldsymbol{\lambda}_i)$ is computed by projecting the odeco polynomial $h={\lambda}_i^x x^4+{\lambda}_i^y y^4 +{\lambda}_i^z z^4$ to the spherical harmonics (SH) basis, where ${\lambda}_i^z$ corresponds to the $z$-axis (so as for $x, y$). This projection is lossless and can be encoded using bands 0, 2, and 4 of SH according to the definition in \cite{green2003sphericalHarmonic}. 
As shown in Fig.~\ref{fig:fz}, this allows us to represent the canonical odeco tensor $\hat{f}(\boldsymbol{\lambda}_i) \in \mathbb{R}^{15}$ as a weighted summation of SH bases (lobes), which can be written as 
\begin{equation}\label{eq:f_canonical}
    \hat{f}(\boldsymbol{\lambda}_i) = \boldsymbol{B}\boldsymbol{\lambda}_i,
\end{equation}
where $\boldsymbol{B}\in \mathbb{R}^{15\times 3}$ is a constant matrix derived from SH bases (band 0, 2, 4) and $\boldsymbol{\lambda}_i\in \mathbb{R}^3$ are axis-aligned stretching ratios. The \emph{major lobe} corresponds to the largest stretching ratio. The full expression is provided in Section 1.2 of Supplementary Material. 

\subsection{Boundary Odeco Tensor}
One often desires the odeco tensor frame field to conform to the domain boundary, i.e., aligning one of its lobes to the surface normal.
We achieve this by rotating the z-axis of an odeco tensor to align with the normal $\boldsymbol{\vec{n}}_i$ of a boundary vertex, and then constrain the orientation parameter to only rotate around $\boldsymbol{\vec{n}}_i$ (equivalently the z-axis). This leads to a simplified odeco tensor expression 
\begin{equation}\label{eq:f_bound}
    f(\boldsymbol{\theta}_i,\boldsymbol{\lambda}_i) = f(\theta_i^z,\boldsymbol{\lambda}_i) = \boldsymbol{R}_i e^{\theta_i^z \mathbf{L}_z} \hat{f}(\boldsymbol{\lambda}_i), \quad \forall i\in \partial\Omega,
\end{equation}
for boundary vertices. We use $\boldsymbol{R}_i$ to denote the constant $\mathbb{R}^{15 \times 15}$ rotation matrix that rotates the z-axis of the canonical frame to align with $\boldsymbol{\vec{n}}_i$, and $\hat{f}$ is the same canonical odeco tensor in Eq.~\ref{eq:f_canonical}. The derivation of $\boldsymbol{R}_i$ is provided in Section 1.3 of Supplementary Material.

\subsection{Visualization}\label{se:vis}
\revise{An odeco tensor can be equivalently represented by either an odeco matrix or an odeco polynomial. Thus, we visualize the odeco tensor by either an odeco polynomial (e.g., Fig.~\ref{fig:fz}) or the cuboid-style (e.g., Fig.~\ref{fig:compare_curvature}) or ellipsoid-style glyph (e.g., Fig.~\ref{fig:dragon}) that is created by the corresponding eigenvalues and eigenvectors. Eigenvalues and eigenvectors are obtained by singular value decomposition (SVD) of the tensor matrix. The volumetric odeco tensor field is also visualized by the integral curves (e.g., Fig.~\ref{fig:teaser}). These integral curves are initially randomly sampled and traced along the eigenvector of the largest eigenvalue.
}

\subsection{Comparison Against Baselines}
\begin{figure}
    \centering
    \includegraphics[width=\linewidth]{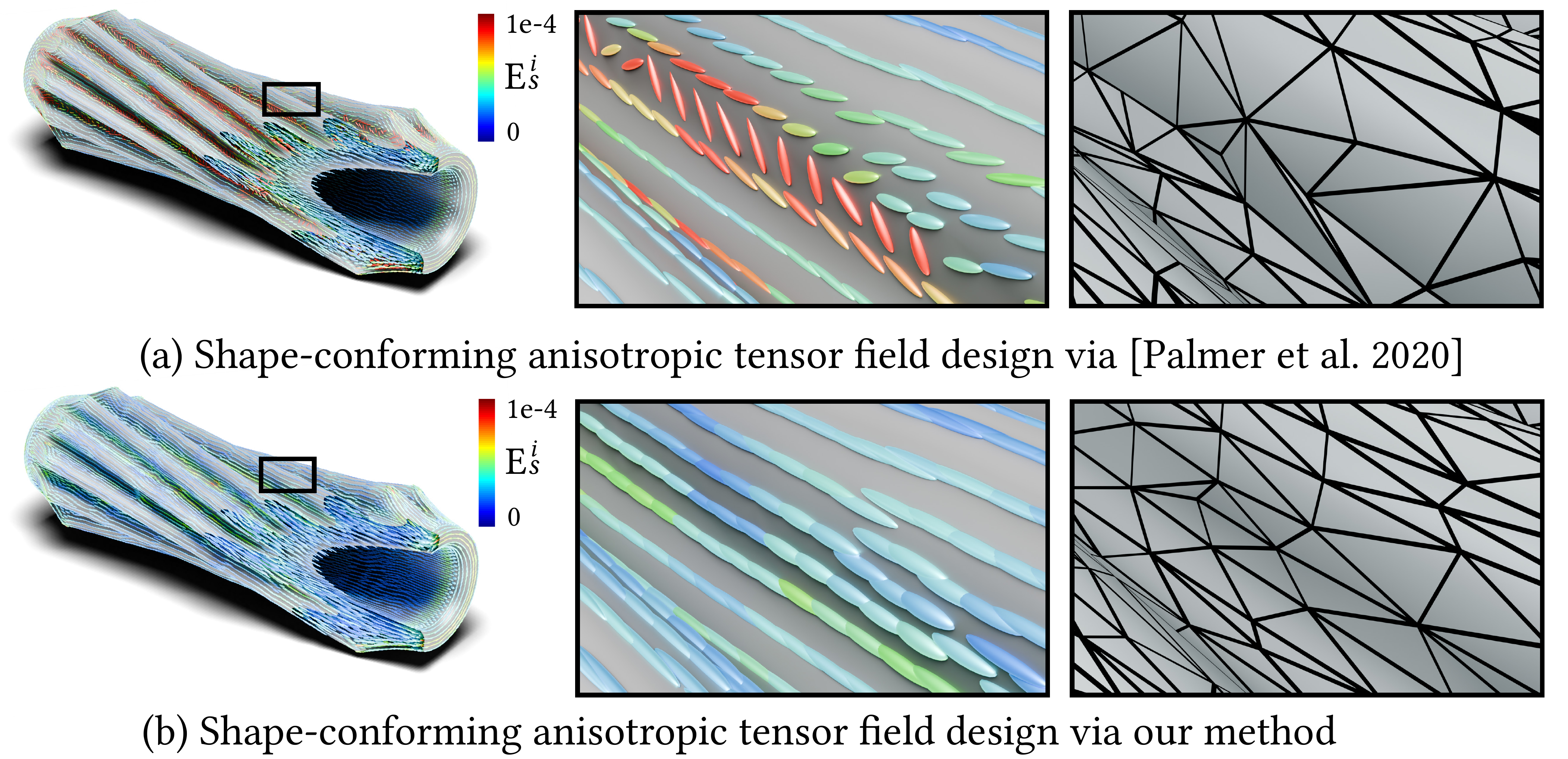}
    \caption{Comparison of shape-conforming anisotropic tensor (frame) field designs. (a) Anisotropic tensor field designed via the method of~\cite{palmer2020algebraic}, guided by both estimated principal curvature magnitudes and directions. (b) Anisotropic tensor field designed by our AOTF method guided only from curvature magnitudes. Our optimal AOTF demonstrates automatic and better shape conformity, a highly desirable characteristic for one important downstream application -- anisotropic meshing.}
    \label{fig:motivation_demo}
\end{figure}

Our odeco tensor representation \cite{robeva2016orthogonal} shares similarities with previous methods for computing \emph{isotropic} tensor fields, such as \cite{palmer2020algebraic, zhang2020featureframe}.  
\revise{In this work, we adopt the same odeco tensor representation that is first used by \citet{palmer2020algebraic}. However, their objective is to compute smoothed tensor fields that produce nearly \emph{isotropic} results, we instead address both smoothed and \emph{anisotropic} tensor field design. Due to the difference in problem formulation, our approach has a different novel objective function, a different novel optimization strategy, and a different set of optimization variables (angles and scaling). We make modifications to \cite{palmer2020algebraic} so that we can compare with their method on anisotropic tensor field design to show the difference.}
As illustrated in Fig.~\ref{fig:motivation_demo}, we show that the way of optimization in \cite{palmer2020algebraic} struggles in getting smooth 3D anisotropic tensor fields since they rely on the prescribed completed tensor. However, the input anisotropic guidance, e.g., estimated curvature tensors, usually has noisy or unstable orientations and stretching ratios for challenging geometric models. This limitation motivates the development of our formulation in Section~\ref{se:Methodology}. 
For the rest of the paper, we still use the same references \cite{palmer2020algebraic}, to indicate their proposed solvers, even though we modify their constraints to take in anisotropic inputs.

\begin{figure*}
    \centering
    \includegraphics[width=\linewidth]{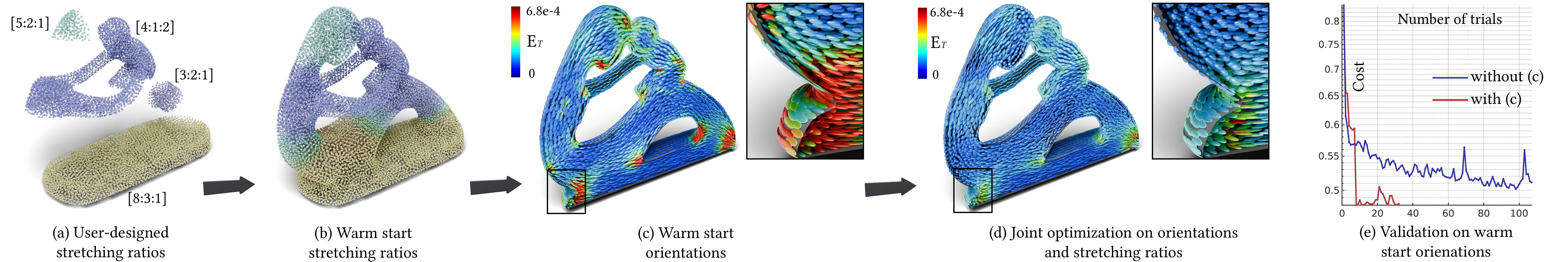}
    \caption{The pipeline of our warm start-based joint optimization of tensor orientations and stretching ratios. (a) Given $\boldsymbol{\Lambda^{In}}$ (user's sparse input) denoted by different colors; (b) Warm start stretching ratios $\boldsymbol{\Lambda}_{warm}$ in volume domain by diffusion; (c) Warm start orientations $\boldsymbol{\Theta}_{warm}$ with coordinate descent method; (d) Joint optimization on orientations and stretching ratios with $\boldsymbol{\Lambda}_{warm}$ and $\boldsymbol{\Theta}_{warm}$ as initial guesses. Step (c) accelerates the procedure of searching better local minima as shown in (e). $E_{T}$ is the total energy of AOTF in Eq.~(\ref{eq:problem}).}
    \label{fig:fertility_global} 

\end{figure*}

\section{Methodology}\label{se:Methodology}
\label{se:smoothness}
Given a tetrahedral mesh with a sparse set of user-specified stretching ratios, our goal is to compute a 3D anisotropic odeco tensor field (AOTF) that is smooth, aligned with domain boundary and user guidance. 

\subsection{Formulation}
\label{se:problem}
To achieve this, we cast it an optimization problem 
\begin{equation}\label{eq:problem}
\begin{aligned}
    \min_{\boldsymbol{\Theta},\boldsymbol{\Lambda}} \; E_{T} &=E_{s}(\boldsymbol{\Theta},\boldsymbol{\Lambda})+ \psi E_{\boldsymbol{\Lambda}}(\boldsymbol{\Lambda}),\\
    E_{s}(\boldsymbol{\Theta},\boldsymbol{\Lambda}) &= \sum_{i,j\in \Omega} w_{ij} ||f(\boldsymbol{\theta}_i,\boldsymbol{\lambda}_i)-f(\boldsymbol{\theta}_j,\boldsymbol{\lambda}_j)||_2^2,\\
    E_{\boldsymbol{\Lambda}}(\boldsymbol{\Lambda})&=\sum_{i\in \Omega} ||\boldsymbol{\lambda}_{i}-\boldsymbol{\lambda}^{In}_{i}||_2^2,
\end{aligned}
\end{equation}
where the total energy $E_T$ consists of a Dirichlet energy term $E_s$ for measuring the smoothness of the field, and $E_{\boldsymbol{\Lambda}}$ denotes a soft penalty term to encourage alignment with user-specified stretching ratios $\boldsymbol{\Lambda}^{In}$ controlled by a weighting term $\psi$ (we set the default value of $\psi = 50$). We provide more details about the user-designed guidance $\boldsymbol{\Lambda}^{In}$ in Section~\ref{se:guidance}.
We follow the method by \cite{crane2019n} to discretize the Dirichlet energy, which leads to $w_{ij}$ being the cotangent weight for the tetrahedral mesh.

As mentioned in Section~\ref{se:algebraic}, the odeco tensor $f(\boldsymbol{\theta}_i,\boldsymbol{\lambda}_i)$ (or in short $f_i$) is controlled by the orientation $\boldsymbol{\theta}_i$ and the stretching $\boldsymbol{\lambda}_i$ parameters. If a vertex $i \in \Omega$ lies on the interior domain, we have the full degrees of freedom to optimize the rotation $\boldsymbol{\theta}_i \in \mathbb{R}^3$ along each axis (see Eq.~(\ref{eq:f_interior})) and the stretching ratios $\boldsymbol{\lambda}_i \in \mathbb{R}^3$. If a vertex $i \in \partial \Omega$ is on the boundary of the domain, in addition to $\boldsymbol{\lambda}_i$, we only optimize the rotation $\theta_i^z$ around the normal direction to encourage boundary alignment (see Eq.~(\ref{eq:f_bound})).
This constraint in the boundary rotation parameter encourages our result to conform to the surface, which is a desirable property for applications in Section~\ref{se:applications}. We defer readers to Section~\ref{se:analysis_shapeconform} for a more detailed and theoretical discussion on shape conformity for curvature alignment and feature preservation. 

\subsection{Optimization}\label{se:Optimization}
Due to the differentiability of our energy, we adopt the L-BFGS~\cite{liu1989limited} to solve Eq.~(\ref{eq:problem}). However, as our energy is non-convex, we empirically discover that naively running L-BFGS from random initializations often leads to bad local minima for complex meshes (see Fig.~\ref{fig:fertility_global} (e)). 
In this section, we thus discuss our strategies to \emph{warm start} the L-BFGS, which leads to a much faster convergence to a better local minimum compared to \emph{cold start}.

\subsubsection{Warm Start $\boldsymbol{\Lambda}$}
Given some user-specified stretching ratios $\boldsymbol{\Lambda}^{In}$, as shown in Fig.~\ref{fig:fertility_global} (a). We warm start the stretching ratios $\boldsymbol{\Lambda}$ by diffusing $\boldsymbol{\Lambda}^{In}$ to the entire volume (see Fig.~\ref{fig:fertility_global} (b)). 
We treat stretching ratios along each axis $\lambda_i^x, \lambda_i^y, \lambda_i^z$ as independent scalar values, and diffuse them with the implicit Euler method. 
We use $\boldsymbol{\Lambda}_{warm}$ to denote the warm started stretching ratios from the diffusion.

\subsubsection{Warm Start $\boldsymbol{\Theta}$}\label{se:warm_start_theta}
Warm starting the orientation field $\boldsymbol{\Theta}$ is less trivial. One could warm start the orientations with the solutions designed for isotropic fields, such as \cite{palmer2020algebraic}. But because the computed orientation completely ignores the anisotropy, warm starting $\boldsymbol{\Theta}$ with the method by \cite{palmer2020algebraic} only leads to incremental improvement on the convergence. 

We instead warm start the orientation field by fixing the (warm started) stretching ratios $\boldsymbol{\Lambda}_{warm}$ and solving for warm started orientations $\boldsymbol{\Theta}_{warm}$ with L-BFGS, inspired by the coordinate descent method (see Fig.~\ref{fig:fertility_global} (c)). Inspired by the Monte Carlo method~\cite{metropolis1949monte,lu2012global}, we iterate between (1) running L-BFGS solely on the orientations and (2) adding perturbations to the current orientations to strive for converging to a good local minimum. 
As this warm start computation only happens on the orientations $\boldsymbol{\Theta}_{warm}$, it is much faster due to a smaller problem size compared to the global L-BFGS (detailed in Section~\ref{se:joint_optimization}), and the result is aware of the ``rough'' anisotropy from $\boldsymbol{\Lambda}_{warm}$. 
However, such a coordinate descent solver may struggle to converge to good solutions (e.g., \cite{BouazizMLKP14}). We experience the same issue if we merely use a coordinate descent solver back and forth between $\boldsymbol{\Lambda}, \boldsymbol{\Theta}$. We thus merely use this as a tool as a fast warm start on the orientations, and still solve $\boldsymbol{\Lambda}, \boldsymbol{\Theta}$ jointly for a better convergence.

\subsubsection{Joint Optimization $\boldsymbol{\Lambda}$ \& $\boldsymbol{\Theta}$}\label{se:joint_optimization}
Given $\boldsymbol{\Lambda}_{warm}, \boldsymbol{\Theta}_{warm}$ from the above two steps, we adopt the same method as Section~\ref{se:warm_start_theta} which iterates between (1) running L-BFGS until convergence and (2) adding parameter perturbations, until the total loss $E_T$ staggers for $5$ consecutive \emph{trials} (see Fig.~\ref{fig:fertility_global} (d)). Note that we use ``trial'' to denote one L-BFGS solve plus one perturbation to differentiate the ``iteration'' count inside each L-BFGS solve.
We adaptively perturb the parameters $\boldsymbol{\theta}_i, \boldsymbol{\lambda}_i$ for each vertex $i$ based on how well it optimizes our energy Eq.~(\ref{eq:problem}). Specifically, we measure the smoothness of each vertex via the same Dirichlet energy, but on the one-ring neighbors $\mathbf{N}_i$ as follows,

\begin{equation}\label{eq:vertexE}
    E_s^i=\frac{1}{2}\sum_{j\in \mathbf{N}_i} w_{ij} ||f(\boldsymbol{\theta}_i,\boldsymbol{\lambda}_i)-f(\boldsymbol{\theta}_j,\boldsymbol{\lambda}_j)||_2^2, \,j\neq i \in \Omega,
\end{equation}
and compute the perturbation amount using

\begin{equation}\label{eq:adap_perturb}
\gamma_i= (E_s^i/max(E_s^i) +1)^2 \, rand(-\epsilon,\epsilon),
\end{equation}
where the magnitude of the random noise $\epsilon$ is set to $0.15$ in all our experiments. Intuitively, this strategy adds bigger perturbations around rough regions, and vice versa. Empirically, this strategy performs particularly well in regions with singularities. The noise bound facilitates jumps within the solution space and consistently lead to smoother results compared cold start the solution throughout our experiments.

\section{Analysis on Shape Conformity}\label{se:analysis_shapeconform}
In this section, we provide the theoretical analysis of the shape conformity in our problem Eq.~(\ref{eq:problem}).
To discuss the shape conformity, we have to particularly analyze the boundary smoothness $E_{s}^{B}(\boldsymbol{\Theta},\boldsymbol{\Lambda}), i,j\in \partial\Omega$, which can be decomposed from $E_{s}$. We will reveal that $E_{s}^{B}(\boldsymbol{\Theta},\boldsymbol{\Lambda})$ is dominant to demonstrate shape conformity if there is a deviation energy term on the boundary. To ease the analysis, the deviation term to user-designed stretching ratios $\boldsymbol{\Lambda}^{In}$ will be simplified as hard constraints in local geometry.

\subsection{Boundary Curvature Alignment}\label{se:smooth_bound}
\begin{figure}[t]
    \centering
\includegraphics[width=\linewidth]{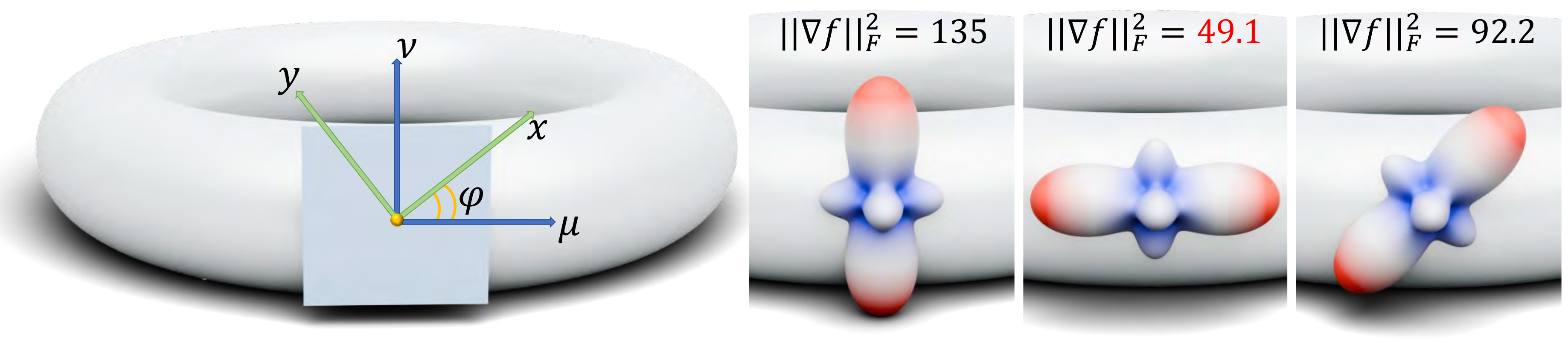}
    \caption{Dirichlet energy of a normal-aligned odeco tensor. In this example, we set the stretching ratios as $\lambda^x_i=2,\lambda^y_i=1,\lambda^z_i=1$. The computed curvature-aligned term of $||\nabla f(\boldsymbol{\theta}_i,\boldsymbol{\lambda}_i)||_2^2$ coincides with Proposition~\ref{prop_curvaalign}.}
    \label{fig:grad_energy}
\end{figure}

We first investigate the integrated $||\nabla f(\boldsymbol{\theta}_i,\boldsymbol{\lambda}_i)||_2^2$ on the surface, which transforms 
$E_{s}^{B}(\Theta,\Lambda)$ to a continuous representation. It measures the amount of local variation of the odeco tensor on a smooth boundary.
\begin{proposition}\label{prop_dirichlet}
Let $f(\boldsymbol{\theta}_i,\boldsymbol{\lambda}_i), i \in \partial\Omega$ be a normal-aligned odeco tensor on a smooth surface $\partial\Omega$ with stretching ratios $({\lambda}^x_i,{\lambda}^y_i,{\lambda}^z_i)$. $f(\boldsymbol{\theta}_i,\boldsymbol{\lambda}_i)$ is re-expressed as $f=\boldsymbol{R}_i e^{\theta_i^z L_z}e^{\phi L_z}\hat{f},$ where $\phi$ is the rotation angle in the tangent plane from the default axis $(x,y)$ to the principal curvature directions $(\mu,\nu)$, as shown in Fig.~\ref{fig:grad_energy}. Thus, we have
\begin{equation}\label{eq:gradiendf}
\begin{aligned}
    ||\nabla f(\boldsymbol{\theta}_i,\boldsymbol{\lambda}_i)||_2^2 = &(cos^2(\phi)g_1(\boldsymbol{\lambda}_i)+sin^2(\phi)g_2(\boldsymbol{\lambda}_i)){K_{max}}^2+\\
    &(sin^2(\phi)g_1(\boldsymbol{\lambda}_i)+cos^2(\phi)g_2(\boldsymbol{\lambda}_i)){K_{min}}^2+\\
    &g_3(\boldsymbol{\lambda}_i)\omega,
\end{aligned}
\end{equation}
where $K_{max}$ and $K_{min}$ are the principal curvatures at vertex $i$. $\omega=(\frac{\partial \theta^z_i}{\partial \mu})^2+(\frac{\partial \theta^z_i}{\partial \nu})^2$ reflects the intrinsic tangential twist around the normal at vertex $i$. And $g_k(\boldsymbol{\lambda}_i)=\frac{64\pi}{315}(4({\lambda}_i^{m_k}-{\lambda}_i^{n_k})^2+({\lambda}_i^{m_k}+{\lambda}_i^{n_k})^2), \; m_k, n_k \in \{\{y,z\},\{x,z\},\{x,y\}\}$ for $k=1,2,3$.
\end{proposition}

The full proof of this proposition is provided in Section 1.4 of Supplementary Material. The smoothness energy $||\nabla f(\boldsymbol{\theta}_i,\boldsymbol{\lambda}_i)||_2^2$ is divided into the extrinsic curvature-aligned term that involves $K_{max}, K_{min}$ and the intrinsic tangential twisting term $g_3(\boldsymbol{\lambda}_i)\omega$. And
$g_i(\boldsymbol{\lambda}_i)$ are free to be reduced if no additional constraints on $\boldsymbol{\lambda}_i$, therefore, minimizing $||\nabla f(\boldsymbol{\theta}_i,\boldsymbol{\lambda}_i)||_2^2$ naturally leads to small and identical stretching ratios. This is why we need user-specified stretching ratios $\Lambda^{In}$ as guidance to introduce anisotropy. If ${\lambda}^x_i = {\lambda}^y_i={\lambda}^z_i$, we have $g_1(\boldsymbol{\lambda}_i) = g_2(\boldsymbol{\lambda}_i) = g_3(\boldsymbol{\lambda}_i)$, which coincides with the result of the octahedral field in ~\cite{zhang2020featureframe}. It is a special case of the odeco tensor field. Then, we analyze the situation of $\boldsymbol{\lambda}_i=\boldsymbol{\lambda}^{In}_i, i\in\partial\Omega$ at local geometry.

\begin{proposition}\label{prop_curvaalign}
When $\boldsymbol{\lambda}_i=\boldsymbol{\lambda}^{In}_i$ with ${\lambda}^x_i \neq {\lambda}^y_i$ and ${\lambda}^z_i \neq \frac{5}{6}({\lambda}^x_i+{\lambda}^y_i)$, minimizing the extrinsic curvature-aligned term in Eq.~(\ref{eq:gradiendf}) requires that the odeco tensor lobes need to be aligned with principal curvature directions. Moreover, when ${\lambda}^z_i < \frac{5}{6}({\lambda}^x_i+{\lambda}^y_i)$, the lobe with a larger stretching ratio must align with the minimum principal curvature direction.
\end{proposition}

The full proof of this proposition and a demo example are provided in Sections 1.5 and 1.7 of Supplementary Material. The case of $\lambda^x_i \neq \lambda^y_i$ and $\lambda^z_i \neq \frac{5}{6}(\lambda^x_i+\lambda^y_i)$ offers shape guidance and allows for automatic curvature alignment when optimizing a normal-aligned AOTF with user-specified guidance $\Lambda^{In}$.
\begin{figure}[t]
    \centering
    \includegraphics[width=\linewidth]{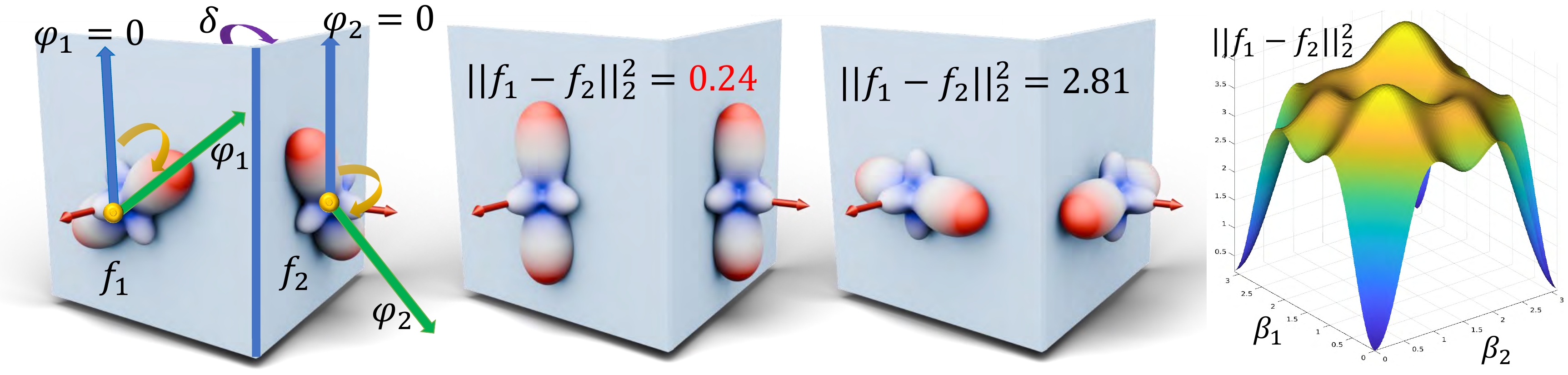}
    \caption{A natural feature preservation of odeco tensors when minimizing $||f_1-f_2||_2^2$. For this example, the stretching ratios and the dihedral angle between two tangent planes are $\lambda^x_i=2,\lambda^y_i=1,\lambda^z_i=1, \delta=0.6\pi$. The plotted distance coincides with Proposition~\ref{prop_nearfeature}.}
    \label{fig:feature_jump}
\end{figure}

\subsection{Boundary feature preservation}\label{se:crease_bound}
Since the normal direction on the sharp feature edges is always ill-defined, the major challenge when applying feature preservation on wrinkled edges is determining the orientation of the odeco tensors. Fortunately, the discretized Dirichlet energy on AOTF usually offers a natural feature preservation along both weak and sharp features in local areas, which is analyzed as follows.

\begin{proposition}\label{prop_nearfeature}
Let $f(\boldsymbol{\theta}_1,\boldsymbol{\lambda}_1^{In}), f(\boldsymbol{\theta}_2,\boldsymbol{\lambda}_2^{In})$ (for short $f_1, f_2$) represent two normal-aligned odeco tensors whose tangent planes are intersected near the shared edges, where $f_1, f_2$ have the same fixed stretching ratios. Let $\varphi_1,\varphi_2$ denote the deviation angles between the lobe with the larger stretching ratio on the tangent plane and the direction of their shared edge, as shown in Fig.~\ref{fig:feature_jump}. The cost of $||f_1-f_2||_2^2$ is minimized by $\varphi_1 = \varphi_2=0 \pm n\pi, n\in \mathbb{Z}$.
\end{proposition}

\begin{figure}[t] 
    \centering
    \includegraphics[width=\linewidth]{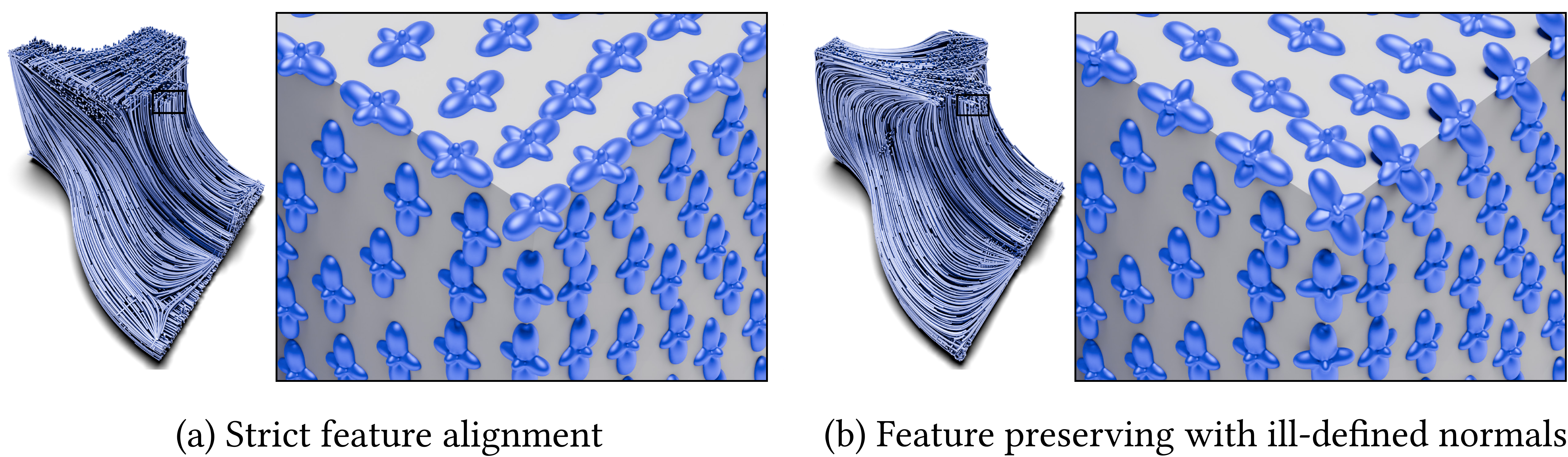} 
    \caption{Optimal orientations on Fandisk model with complex sharp features: (a) applying strict feature alignment constraints and (b) applying original ill-defined normal alignment constraints on the complicated sharp feature regions, respectively.}
    \label{fig:feature_aligned_demo}
\end{figure}

The full proof of this proposition is provided in Section 1.6 of  Supplementary Material. Note that, this proposition is applicable to any pair of odeco tensors with non-parallel tangent planes. Now we analyze why the smoothness energy $E_{s}^{B}(\Theta,\Lambda)$ usually leads to a natural feature preservation. We focus on a local area around the sharp feature edges (without complex sharp feature corners for simplicity).
For a vertex $k$ on the sharp feature edges, its vertex-wise energy is dominated by $\sum (w_{ik}||f_i-f_k||_2^2+w_{jk}||f_j-f_k||_2^2)$, where $f_i,f_j$ are its neighbors on each side of the sharp feature. $f_k$ is adjustable between $f_i$ and $f_j$ with normal constraint, therefore, minimizing $w_{ik}||f_i-f_k||_2^2+w_{jk}||f_j-f_k||_2^2$ subject to minimizing $||f_i-f_j||_2^2$. As proved in Proposition~\ref{prop_nearfeature}, we conclude that minimizing local energy of $E_{s}^{B}(\Theta,\Lambda)$ leads to a natural feature preservation, except for areas of complex features or other constraints, as shown in Fig.~\ref{fig:feature_aligned_demo} (b).

\textit{Strict Feature Alignment}. 
Although $E_{s}^{B}(\Theta,\Lambda)$ can offer feature preservation, we introduce a strategy to achieve the exact alignment to sharp features for challenging scenarios (such as in Fig.~\ref{fig:feature_aligned_demo} (a)). Thus, the sharp feature edges have to be extracted in advance. To resolve the ill-defined normal of the vertices on the feature edges, a natural choice for achieving exact feature alignment is to enforce the major lobes fixed along the sharp feature edges, as we have done for the normal alignment. For corner points which typically are singularities, we arbitrarily assign one of its feature edge directions since their stretching ratios will be further optimized.

\section{Experimental Results and Evaluations}\label{se:evaluations}
\subsection{Implementation Details}
\label{se:implementation}
We implement and evaluate our algorithms on a PC with a 3.6GHz Core CPU, 32GB RAM, and Windows 11. We test our algorithms on various 3D models from the Thingi10k dataset~\cite{zhou2016mesh}, with varying resolutions, complicated geometry, and challenging sharp or weak features. The input tetrahedral meshes are generated by TetGen~\cite{hang2015tetgen} from the surface meshes. We list the detailed statistics, including model information, the number of tetrahedral mesh vertices and elements, and computational timings of all the experiments in Table~\ref{tab:statistics}. The anisotropic stretching ratios for the majority of our examples are within the range of $[1,50]$ based on the application needs. Larger stretching ratios can also be accommodated if required. 
Our L-BFGS optimization is implemented by \textit{fmincon} function in MATLAB. L-BFGS is terminated until $\Delta E_T/E_T<1e^{-3}$ for all trials except the last one with $\Delta E_T/E_T<1e^{-8}$. A single trial 
is usually finished in $2$ minutes without particular parallel acceleration. Most of our examples take $10 \sim 20$ trials to get a better local minimum. \textit{The source code of our framework and data will be publicly released after paper acceptance.}
\begin{figure*}[h]
    \centering
    \includegraphics[width=\linewidth]{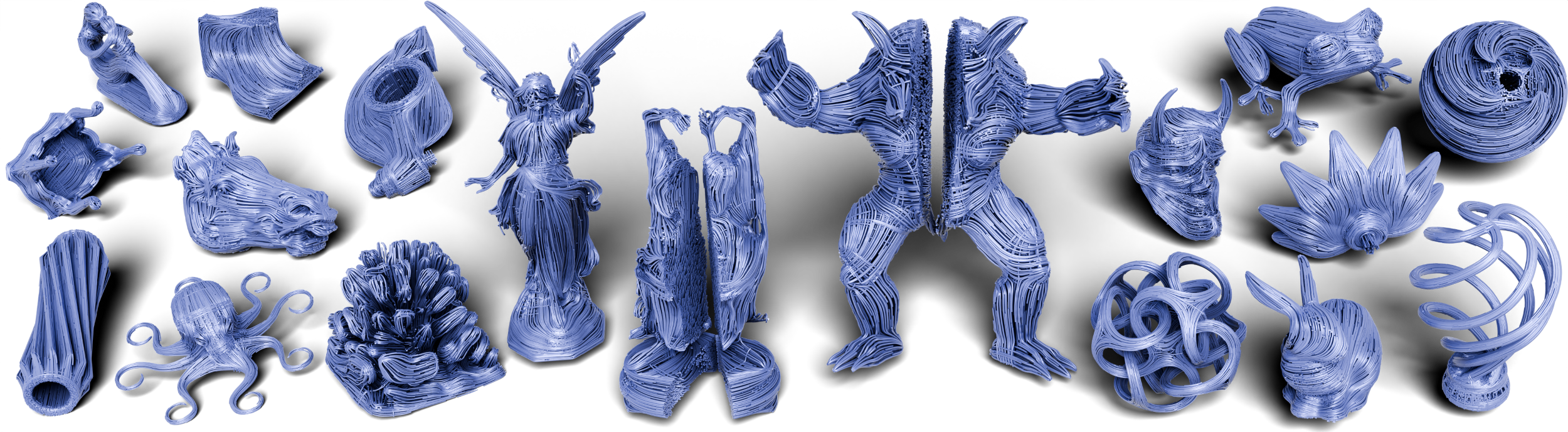}
    \caption{A gallery of our shape-conforming AOTFs on complicated models. The visualization of integral curves are traced along the major lobes of tensors.} 
    \label{fig:more_results}
\end{figure*}

\subsection{User's Design Guidance}
\label{se:guidance}
\begin{figure*}[h]
    \centering
    \includegraphics[width=\linewidth]{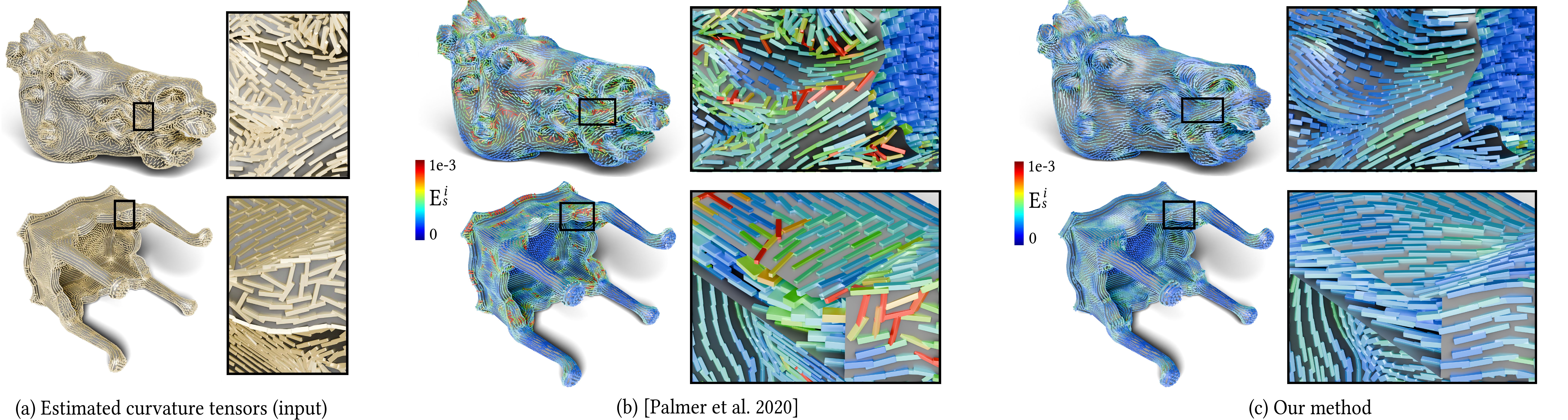}
    \caption{Comparison between our method and the method of~\cite{palmer2020algebraic} in shape-conforming tensor field designs for complicated 3D models. \revise{All tensor elements are illustrated as cuboid-style glyphs to better show the feature preservation.} (a) Estimated curvature tensor fields as inputs; (b) Results of~\cite{palmer2020algebraic} using (a) as the constraints (both curvature magnitudes and directions); (c) Our results using only the curvature magnitudes as the constraints. The zoom-in views depict the boundary and interior of the tensor fields. The comparison demonstrates that we provides a more flexible approach to introducing shape-conforming anisotropy, without necessarily relying on boundary noisy curvature tensors as guidance. $E^i_{s}$ is the vertex-wise smoothness energy.}
    \label{fig:compare_curvature}    
\end{figure*}

\begin{figure*}
    \centering
    \includegraphics[width=\linewidth]{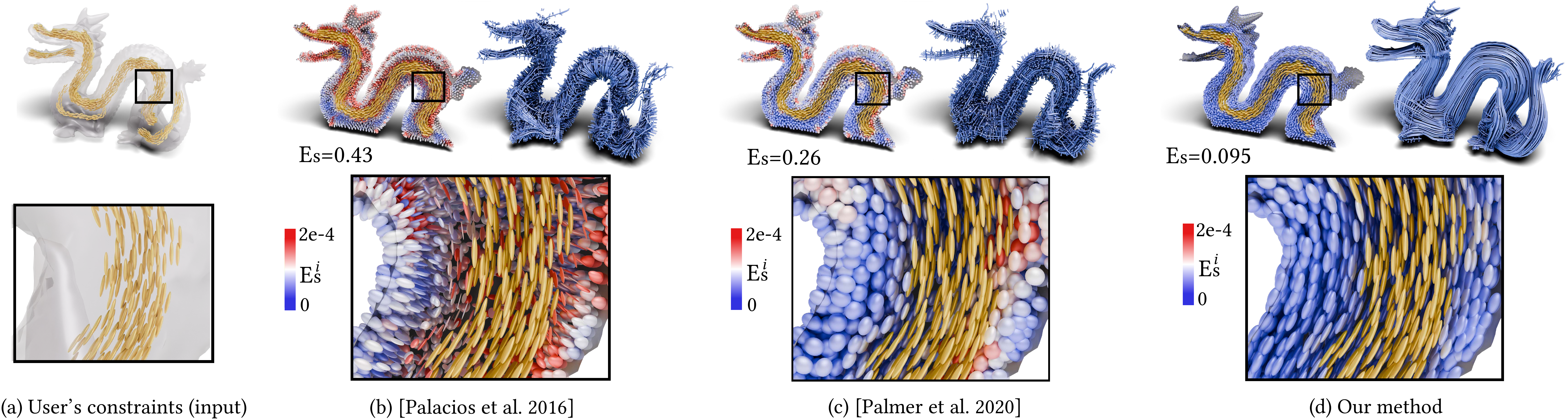}
    \caption{Comparison between our method and the state-of-the-art in smooth tensor field designs from user-specified interior constraints for Dragon model. (a) The yellow ellipsoid-style glyphs indicate the user-specified tensors, which are not to be altered during the design process; (b) Results of~\cite{palacios2016tensor}; (c) Results of~\cite{palmer2020algebraic}; (d) Our results. The clipped zoom-in views depict the interior of the tensor fields. The visualization of integral curves are traced along the major lobes of tensors. The comparison demonstrates that our method generates higher smoothness of 3D tensor fields under user's constraints. $E^i_{s}$ and $E_{s}$ are the vertex-wise and total smoothness energies, respectively. We have the same smoothness energy form of $E_{s}$ as the method~\cite{palmer2020algebraic}.}
    \label{fig:dragon}
\end{figure*}

In this work, we are able to handle different kinds of user's design guidance in the input for our 3D tensor design. The detailed description is given as follows: 
\begin{itemize}
    \item Stretching ratio-based guidance: User-specified stretching ratios at the region of interest in the volume domain as the design guidance are processed by the Blender software, such as Figs.~\ref{fig:fertility_global},~\ref{fig:feature_aligned_demo},~\ref{fig:fab_hand}. 
    \item Tensor-based guidance: User-specified tensors (both stretching ratios and orientations) at the region of interest in the volume domain as constraints are designed by the interactive tensor design framework in~\cite{palacios2016tensor}, such as Figs.~\ref{fig:teaser},~\ref{fig:dragon},~\ref{fig:fab_hand}. The specified tensors can be either degenerate or non-degenerate, distributed at specific points or along curves. More details are introduced in~\cite{palacios2016tensor}. 
    \item Curvature-based guidance: To simplify some design process, the stretching ratios on the boundary can be provided based on the estimated discrete curvatures, such as Figs.~\ref{fig:teaser},~\ref{fig:motivation_demo},~\ref{fig:more_results},~\ref{fig:compare_curvature},~\ref{fig:meshing_sur},~\ref{fig:meshing_vol}. The principal curvature magnitudes $\boldsymbol{K}_{max}, \boldsymbol{K}_{min}$ and corresponding directions are calculated using Libigl~\cite{libigl}. The stretching ratios obtained from the raw estimated curvatures are pre-processed as ${\lambda^x_i}=abs(K_{max}^i/K_{min}^i), {\lambda^y_i}=1$. ${\lambda^x_i}$ is clamped to a reasonable range $[1,50]$. ${\lambda^z_i}$ that corresponds to the normal is typically treated as a free variable. However, if it needs to be explicitly specified, we ensure that the situation where ${\lambda^z_i} = \frac{5}{6}({\lambda^x_i}+{\lambda^y_i})$ is avoided, according to Proposition~\ref{prop_curvaalign}. 
    \item Field-based guidance: User can load any raw (pre-computed) volumetric tensor field as an initialization for further process under our framework, such as the stress-simulated tensor fields are used in our microstructure fabrication in Figs.~\ref{fig:fab_helmet},~\ref{fig:fab_block}.
\end{itemize}

\begin{table}
\caption{Statistics of our experiments. Note: \# denotes the number of elements. $\hat{T}_{\boldsymbol{\Lambda}_{warm}}$ denotes the time (s) for the warm start $\boldsymbol{\Lambda}$; 
$\hat{T}_{\boldsymbol{\Theta}_{warm}}$ denotes the averaged time (s) for a single trial during the warm start $\boldsymbol{\Theta}$;
\#$T_{\boldsymbol{\Theta}_{warm}}$ denotes the number of trials for warm start $\boldsymbol{\Theta}$;
$\hat{T}_{\boldsymbol{\Theta},\boldsymbol{\Lambda}}$ denotes the averaged time (s) for a single trial during the joint optimization; \#$T_{\boldsymbol{\Theta},\boldsymbol{\Lambda}}$ denotes the number of trials for the joint optimization;  
$\hat{T}_{total}$ denotes the total computational time (s).}
\label{tab:statistics}
\begin{center}
\begin{adjustbox}{width=\linewidth,center}
    \begin{tabular}{llllllllll}
    \toprule
     Models &  \#$\mathcal{V}$ & \#$\mathcal{T}$&$\lambda_{max}$&$\hat{T}_{\boldsymbol{\Lambda}_{warm}}$&$\hat{T}_{\boldsymbol{\Theta}_{warm}}$ &\#$T_{\boldsymbol{\Theta}_{warm}}$& $\hat{T}_{\boldsymbol{\Theta},\boldsymbol{\Lambda}}$ &\#$T_{\boldsymbol{\Theta},\boldsymbol{\Lambda}}$&$\hat{T}_{total}$\\
    \midrule
    Fig.~\ref{fig:teaser}~Lucy& 69,834 &320,339&50&0.23&43&11&67&5&808\\
    Fig.~\ref{fig:motivation_demo}~Vase &60,824&283,473&30&0.19&41&8&63&5&643\\
    Fig.~\ref{fig:fertility_global}~Fertility &23,263&113,179&8&0.07&18&12&37&6&438\\
    Fig.~\ref{fig:more_results}~Fandisk &26,554&120,800&4&0.12&19&14&41&5&471\\
    Fig.~\ref{fig:more_results}~Rockerarm &32,758&167,030&15&0.17&24&8&39&7&465\\
      Fig.~\ref{fig:more_results}~Budda &68,677&343,528&50&0.23&51&15&72&6&1,197\\
       Fig.~\ref{fig:more_results}~Armadillo &79,398&387,624&50&0.29&69&9&84&4&957\\
        Fig.~\ref{fig:more_results}~Bunny &26,779&180,940&4&0.11&18&8&24&5&264\\
         Fig.~\ref{fig:more_results}~Frog &49,965&260,425&20&0.2&33&8&47&5&499\\
         Fig.~\ref{fig:more_results}~Octopus &10,523&46,803&15&0.04&11&13&21&7&290\\
          Fig.~\ref{fig:compare_curvature}~Top &43,657&225,844&50&0.16&28&8&38&5&414\\
          Fig.~\ref{fig:compare_curvature}~Bottom &26,554&120,800&50&0.10&17&6&26&5&232\\
          Fig.~\ref{fig:dragon}~Dragon &31,062&142,340&10&0.11&24&9&34&6&420\\
          Fig.~\ref{fig:meshing_sur}~Top &42,261&223,190&20&0.18&31&7&52&5&477\\
          Fig.~\ref{fig:meshing_sur}~Bottom &56,168&241,658&20&0.20&41&6&67&5&581\\
          Fig.~\ref{fig:fab_helmet}~Helmet &71,346&338,364&50&0.39&70&9&91&5&1,085\\
           Fig.~\ref{fig:fab_block}~Block &51,250&272,418&50&0.20&37&11&56&5&681\\
           Fig.~\ref{fig:fab_hand}~Hand &58,388&312,427&8&0.23&41&5&73&5&570\\
    \bottomrule
    \end{tabular}
\end{adjustbox}
\end{center}
\end{table}

\subsection{Evaluation Metrics}
Besides the qualitative visualization measurement for field smoothness and shape conformity (e.g., curvature alignment and feature alignment), the main quantitative evaluation metrics for the tensor field results are the overall smoothness $E_s$ and its vertex-wise version $E_s^i$, where $E_s$ is also a classic Dirichlet energy form and $\sum_{i \in \Omega} E_s^i=E_s$. The energy values are normalized by dividing with the number of input tetrahedral mesh vertices to ensure generality across different models and resolutions. It is noted that the size of each tensor (represented by ellipsoid-style glyphs) is accommodated for visualization while preserving its original stretching ratios. We also provide the visualization of integral curves, which are traced along the major lobes of tensors. Some additional evaluation metrics for the specific applications are provided in Section~\ref{se:applications}. The detailed computational times of the optimization are given in Table~\ref{tab:statistics}, including the computational time for the warm start of the field's stretching ratios $\boldsymbol{\Lambda}$: ${T}_{\boldsymbol{\Lambda}_{warm}}$, the averaged computational time for a single trial during the warm start of the field's orientations $\boldsymbol{\Theta}$: $\hat{T}_{\boldsymbol{\Theta}_{warm}}$, the averaged computational time for a single trial during the joint optimization of the field's stretching ratios and orientations: $\hat{T}_{\boldsymbol{\Theta},\boldsymbol{\Lambda}}$, and the total computational time: $\hat{T}_{total}$. It is noted that the diffusion computation for $\boldsymbol{\Lambda}_{warm}$ is very efficient and ${T}_{\boldsymbol{\Lambda}_{warm}}$ is quite small compared with other steps' timings, which need to run several trials of L-BFGS.

\begin{figure}
    \centering
    \includegraphics[width=\linewidth]{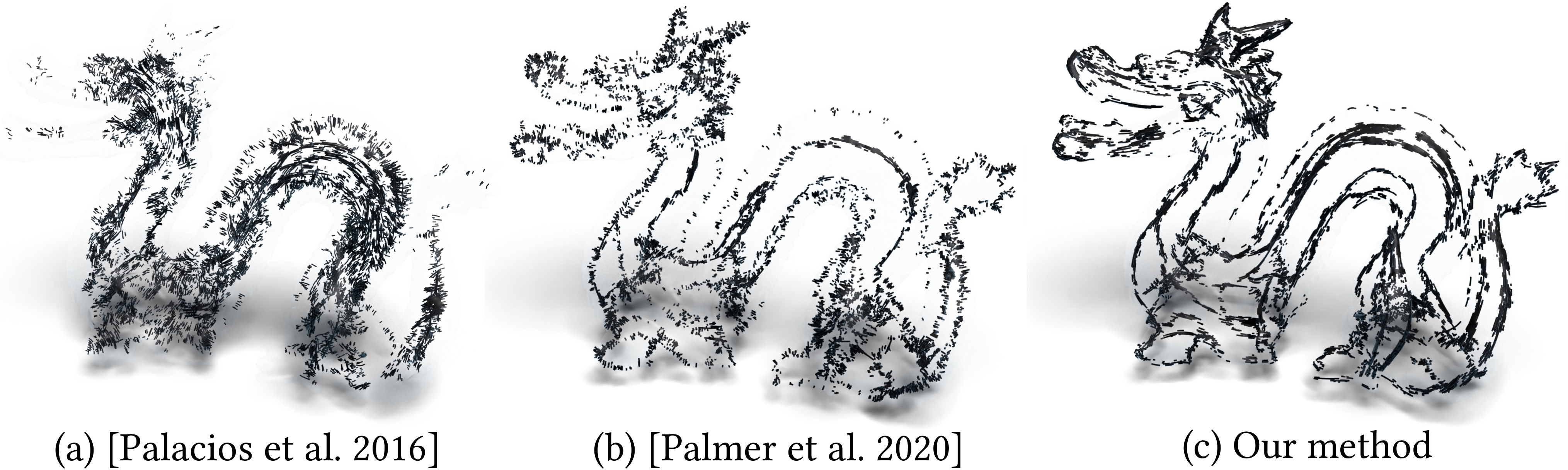}
    \caption{\revise{Additional illustration for the singularity graphs of the results in Fig.~\ref{fig:dragon}. The singular elements are firstly detected by finding the holonomy of the field, which is calculated according to the method of \cite{palmer2020algebraic}. Then, singularity graphs are traced and depicted as integral curves in Section~\ref{se:vis}.}}
    \label{fig:dragon_singular}
\end{figure}

\subsection{Evaluations and Comparisons}
Fig.~\ref{fig:more_results} demonstrates a gallery of our optimal shape-conforming AOTF results. It validates our method on challenging models with complex shape variations and features from the Thingi10k dataset~\cite{zhou2016mesh}. For each model, the only input is the boundary anisotropy guidance coming from the curvature magnitudes (stretching ratios), and the normal alignment and strict feature alignment are automatically applied. The depicted integral curves are initially sampled and traced along the major lobes of tensors. These integral curves effectively align with the local geometry, while maintaining the smoothness with the automatic balance in regions with intricate details and features. In the following, we compare our method with the state-of-the-art in volumetric tensor designs by using odeco representation~\cite{palmer2020algebraic} and symmetric matrix representation~\cite{palacios2016tensor}. 

In Fig.~\ref{fig:compare_curvature}, we compare with~\cite{palmer2020algebraic} in shape-conforming tensor field designs for complicated 3D models. The common constraints are normal alignments in both methods. We have the same smoothness energy form using odeco $E_{s}$ with the method of \cite{palmer2020algebraic}. The results of \cite{palmer2020algebraic} are implemented by their source code. We apply boundary curvature tensor (both magnitudes and directions) as hard constraints in their mMBO-based method in~\cite{palmer2020algebraic}. The boundary odeco tensors are not altered when solving their heat linear system. The main reason of assigning specified curvature tensors is that their orientations and stretching ratios are unified and embedded together in their formulation, which lacks the flexibility of controlling them independently. These hard constraints ensure the desired anisotropy is kept on the boundary; however, the input raw curvature tensor field is always noisy, which detriments the overall field smoothness. However, in our method the orientations can be automatically optimized while taking curvature magnitudes as the only constraints. Note that, we also make the input magnitudes as hard constraints only for comparison. Our optimization framework provides a more flexible approach to introducing shape-conforming anisotropy, without necessarily relying on boundary noisy curvature tensors as guidance. 

In Fig.~\ref{fig:dragon}, we compare with~\cite{palacios2016tensor} and \cite{palmer2020algebraic} in smooth tensor field designs from user-specified constraints. To make a fair comparison, we only apply normal alignment and user-specified interior constraints, which are used by all methods. The results of \cite{palacios2016tensor} are obtained from their developed framework. The results of \cite{palmer2020algebraic} are implemented by their source code. 
Both \cite{palacios2016tensor} and \cite{palmer2020algebraic} have limited freedom on the boundary, which curbs the boundary odeco tensors. A fundamental reason for the artifacts in \cite{palacios2016tensor} is the way of computational operations on tensor matrix element. \cite{palmer2020algebraic} improves the smoothness but still shows a weaker transition near the boundary. The main reason is their un-normalized basis when denoting by a lower-dimensional vector, as well as the projection operation that limits the flexibility of stretching ratios. However, our method outperforms both of them in terms of the lower smoothness energy. Visually, our results achieve a much smoother transition between the user-specified hard constraints and the normal-aligned boundary tensors. \revise{In addition, we demonstrate the corresponding singularity graphs in Fig.~\ref{fig:dragon_singular}.} 

\begin{figure*}
    \centering
    \includegraphics[width=0.96\linewidth]{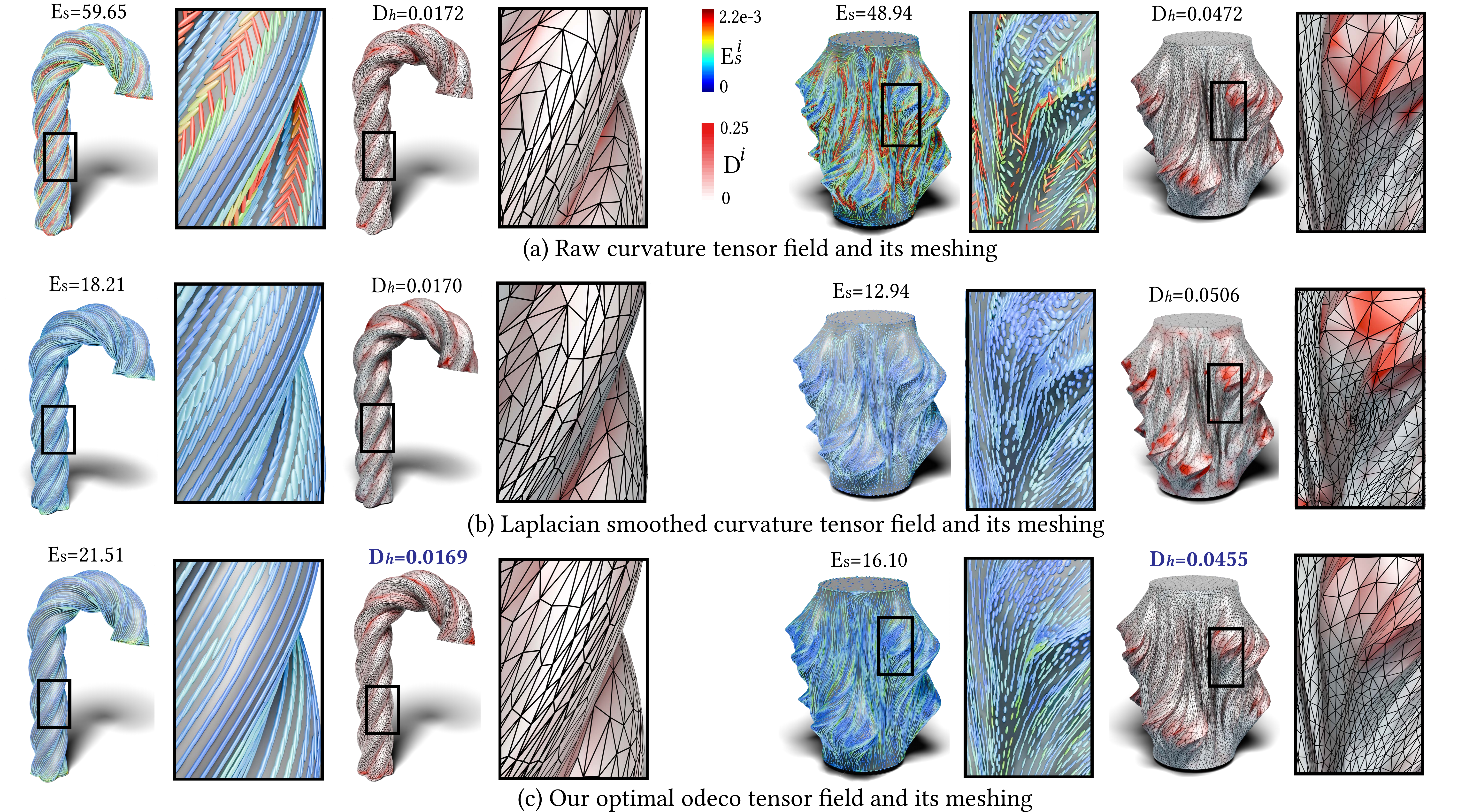}
    \caption{Comparison on feature-sensitive anisotropic triangular meshing~\cite{li2024nasm} results generated from different curvature-based tensor fields. \revise{(a) Raw estimated curvature tensor fields and corresponding meshes with zoom-in views; (b) Smoothed curvature tensor fields via Laplacian method and corresponding meshes with zoom-in views; (c) Our optimal odeco tensor fields and corresponding meshes with zoom-in views. The comparison demonstrates that our optimal odeco tensor fields enable a higher fidelity and better quality meshing result with a smaller Hausdorff distance. } $E^i_{s}$ and $E_{s}$ are the vertex-wise and total smoothness energies, respectively. $D^i$ is vertex-wise minimum distance and $D_{h}$ is the Hausdorff distance between generated mesh and input mesh.}
    \label{fig:meshing_sur}

\end{figure*}

\begin{figure*}
    \centering
    \includegraphics[width=\linewidth]{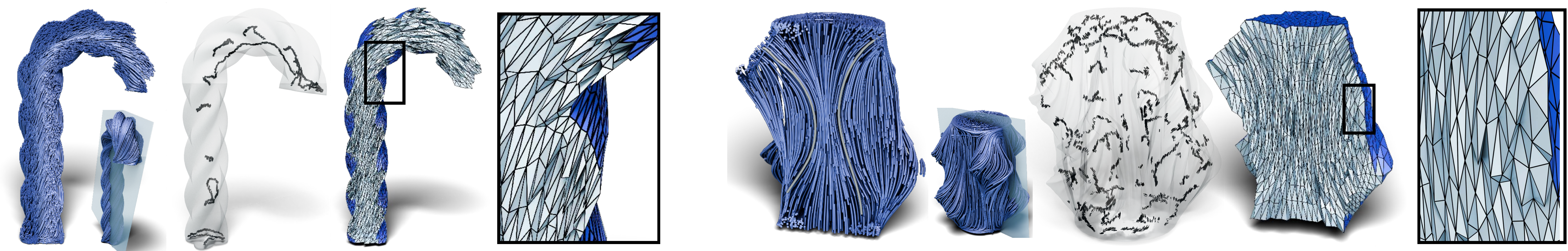}
    \caption{Anisotropic tetrahedral meshing [Zhong et al. 2018] results by using our optimal volumetric odeco tensor fields. \revise{The singularity graphs of the tensor fields' interior are calculated and obtained according to the method of \cite{palmer2020algebraic}. Due to the unstructured nature of tetrahedral meshing, the singularity graphs do not affect the mesh quality.} The tensor field in the model at right shows the user-specified interior constraints depicted as gray curves. The visualization of integral curves are traced along the major lobes of tensors.}
    \label{fig:meshing_vol}

\end{figure*}

\section{Applications}
\label{se:applications}

In addition to the evaluations in the previous section, we introduce some downstream applications of our proposed AOTFs on anisotropic surface and volume meshing, fabrication of anisotropic microstructures, and elastic material designs. 

\subsection{Anisotropic Surface and Volume Meshing}
Anisotropic meshes have great importance in shape approximations and numerical methods to solve partial differential equations. The anisotropic meshes outperform isotropic meshes in scenarios where directional features or variations in the geometry need to be captured effectively. They improve the accuracy of the solution and decrease the computational cost~\cite{alliez2003anisotropic,narain2012anisotropic,huang2005metric}. \revise{The result quality of the current anisotropic meshing methods mainly depend on the input tensor fields~\cite{zhong2018computing}. Our optimal AOTFs offer good quality meshing results with high fidelity and smoothness, primarily because the stretching ratios are preserved rather than smoothed while the orientations are automatically optimized and smoothed, as observed from the comparison with field smoothing in Fig.~\ref{fig:meshing_sur}.}

In Fig.~\ref{fig:meshing_sur}, we demonstrate the performance of our optimal AOTFs on the anisotropic surface and volume meshing. To demonstrate the strength of our shape conformity in our tensor field design, we compare the anisotropic meshes with the curvature tensor fields and our optimal AOTFs. The anisotropic triangular meshing is implemented by the state-of-the-art feature-sensitive anisotropic triangular meshing method in~\cite{li2024nasm}, which is capable of keeping weak features. We use the same number of vertices in the output anisotropic meshes for comparison. We compute the Hausdorff distance $D_h$ between the dense input mesh and the generated anisotropic mesh to measure the ability to express complex local geometric variations of the given shapes. Compared with the estimated raw curvature tensor field (obtained as discussed in Section~\ref{se:guidance}) as input, our optimal AOTFs offer higher fidelity and smoother meshing results. \revise{Compared with the post-smoothed curvature tensor field by Laplacian smoothing used in~\cite{palacios2016tensor}, although their smoothness is improved, the original curvature stretching ratios and boundary geometric features are not well maintained. Furthermore, over-smoothed tensor fields with smaller $E_s$ in Fig.~\ref{fig:meshing_sur} (b) may cause a worse fidelity with a larger Hausdorff distance $D_h$}. The output anisotropic meshes based on our optimal fields can more accurately capture the details and achieve the smaller Hausdorff distance $D_h$ compared with the meshes generated based on the estimated curvature tensor fields. 

Furthermore, we further demonstrate our applications on anisotropic tetrahedral meshing. For instance, we apply our optimal volumetric AOTFs to~\cite{zhong2018computing}. Fig.~\ref{fig:meshing_vol} shows that our AOTFs take user-specified tensors as constraints and curvature magnitudes as guidance to obtain the tetrahedral meshes with both shape conformity and user-desired orientations and stretching ratios. Since current anisotropic tetrahedral meshing methods completely depend on the guidance of the input tensor metrics, our method provides a promising and high-quality input for them. \revise{Note that, the singularity graphs may look complex since we try to keep the large stretching ratios on the boundary. However, due to the unstructured nature of tetrahedral meshing, the singularity graphs do not affect the mesh quality.}

\subsection{Fabrication of Anisotropic Microstructures}
Recent advancements in the design of volumetric anisotropic microstructures, such as stress-aligned anisotropic truss structure~\cite{arora2019volumetric}, anisotropic elastic microstructures~\cite{fabrication2017martinez,martinez2018polyhedral} and orientable structures with anisotropy~\cite{Orientable2020}, rely on tensor fields as input to enable the creation of anisotropic behaviors. \revise{ Previous tensor design methods introduce anisotropy by requiring both user-specified orientations and stretching ratios as input, which demands extensive manual design effort for complex shapes. In contrast, our method automatically optimizes the orientations while only referring to the specified stretching ratios. As a result, the optimized AOTFs naturally conform to surface curvatures and preserve geometric features, leading to improved boundary layers in the generated microstructures.} 

\textit{Tensor Field Smoothing}.
Numerically simulated tensor fields, such as stress-simulated and fluid-simulated, are usually extremely noisy and lack shape conformity. Thus, our method shows the capability of not only increasing the overall smoothness but also balancing the fidelity with the inputs. Similar to Eq.~(\ref{eq:problem}), the tensor field smoothing problem is reformulated as follows:
\begin{equation}\label{eq:Smoothing_problem}
\begin{aligned}
    \min_{\boldsymbol{\Theta},\boldsymbol{\Lambda}} \;& E_{s}(\boldsymbol{\Theta},\boldsymbol{\Lambda})+ \kappa E_{dis},\\
\end{aligned}
\end{equation}
where
\begin{equation}\label{eq:E_dis}
\begin{aligned}
    &E_{dis}=\sum_{i\in \Omega} ||f(\boldsymbol{\theta}_i,\boldsymbol{\lambda}_i)-f^{In}_{i}||_2^2,
\end{aligned}
\end{equation}
where $E_{dis}$ measures the difference between the input tensor field $f^{In}$ and the generated smoothed tensor field $f(\boldsymbol{\theta}_i,\boldsymbol{\lambda}_i)$. $\kappa$ is set as 2 in our experiments. Furthermore, our method has the ability to incorporate normal alignment and feature alignment, but it may lead to a higher value of $E_{dis}$. This is because the input tensor field may have the conflict with the local geometric features. We only run a single trial of L-BFGS for this problem, and the initializations are obtained from the input tensor field.
\begin{figure*}[h]
    \centering
    \includegraphics[width=\linewidth]{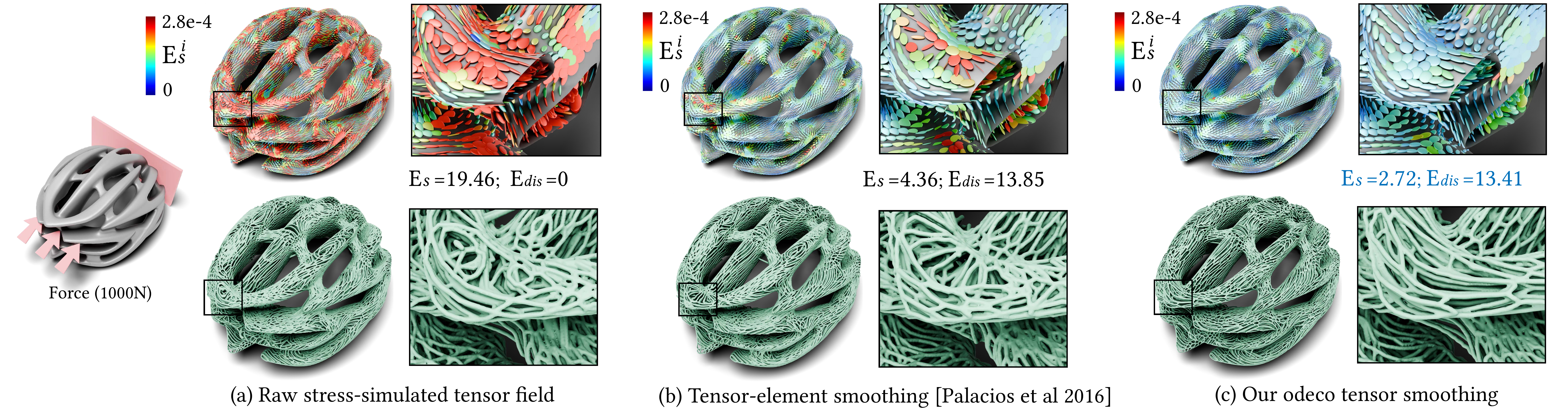}
    \caption{Fabrication of 3D open-cell microstructures for Helmet model from different stress tensor fields. (a) Raw (noisy) stress-simulated tensor field results in unstable microstructures with artifacts; (b) A direct tensor-element smoothing approach~\cite{palacios2016tensor} produces a reasonable microstructure result; (c) Our method produces much smoother (i.e., with a smaller $E_{s}$) odeco tensor fields and the corresponding microstructures while maintaining higher fidelity (i.e., with a smaller $E_{dis}$) to the given input tensor field. $E^i_{s}$ and $E_{s}$ are the vertex-wise and total smoothness energies, and $E_{dis}$ is the total fidelity error.}
    \label{fig:fab_helmet}
\end{figure*}

\begin{figure*}[h]
    \centering
    \includegraphics[width=\linewidth]{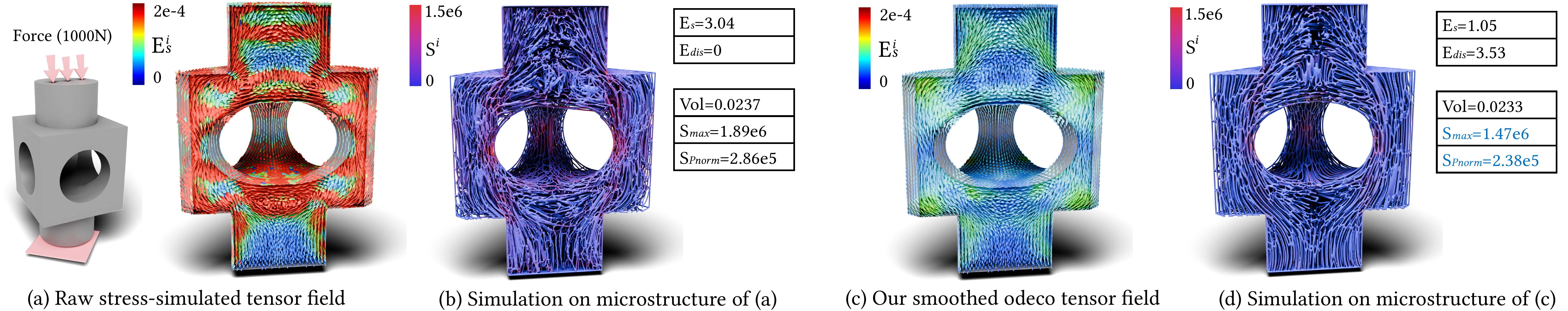}
    \caption{Simulation of 3D open-cell microstructure fabrication for Block model with different stress tensor fields. (a) Raw (noisy) stress-simulated tensor field; (b) Fabrication and simulation results based on (a), which generate unstable microstructures with artifacts as well as a poor static stress simulation; (c) Our smooth odeco tensor fields (i.e., with a smaller $E_{s}$). $E^i_{s}$ and $E_{s}$ are the vertex-wise and total smoothness energies. $E_{dis}$ is the fidelity error; (d) Our method produces higher quality (much smoother) microstructures and better static stress simulation result (i.e., with a smaller $S_{max}$ and $S_{P_{norm}}$). $S^i$ is the vertex-wise Von Mises stress. $S_{max}$ is the maximum Von Mises stress and $S_{P_{norm}}$ is the $P=6$ norm of Von Mises stress. $Vol$ is the total volume of the microstructures.}
    \label{fig:fab_block}
\end{figure*}

\textit{Microstructure Generation}.
In this task, our implementation is built on top of the ideas in~\cite{fabrication2017martinez}, leveraging 3D anisotropic Voronoi diagrams for microstructure generation. The parameters of our open-cell microstructures are anisotropy (orientation and stretching ratios), density (number of cells), and beam thickness. Additionally, they are fabricable and support-free when using SLS-based 3D printers. We provide the designed odeco tensor field as input for anisotropy. The 3D anisotropic Voronoi tessellation is created by the method in~\cite{zhong2018computing}. More details about the microstructure generation can refer from~\cite{fabrication2017martinez}.

In Fig.~\ref{fig:fab_helmet}, we compare our method and the method in~\cite{palacios2016tensor} on smoothing stress-simulated noisy tensor fields and the corresponding microstructures. The raw stress tensor field, which is simulated on the solid object under the force of $1,000N$. The stress simulations on this solid object as well as our microstructures are implemented by the Range software~\cite{range}. We have verified their accuracy by reproducing the same experiments on other commercial software, such as Matlab and Abaqus. The raw stress tensor field usually exhibits significant noise, with stretching ratios ranging from $1$ to $10^6$. For feasibility and practicality, we first apply a logarithmic transformation to map the original ratios to the interval $[1,50]$. One straightforward way of smoothing involves applying Laplacian smoothing directly to the elements in the tensor matrix~\cite{palacios2016tensor}. However, this approach often leads to unreliable results, especially in the area of singularities. The smoothness and fidelity of their method are $E_{s}=4.36$ and $E_{dis}=13.85$. Instead, our method produces much smoother (i.e., with a smaller $E_{s}=2.72$) odeco tensor fields and more stable microstructures while maintaining higher fidelity (i.e., with a smaller $E_{dis}=13.41$) to the given input tensor fields.

In Fig.~\ref{fig:fab_block}, we demonstrate the effectiveness of stress-aligned microstructures generated by our smoothed odeco tensor field. Two microstructures are generated according to the raw simulated stress tensor field and our smoothed odeco tensor field. These two microstructures have almost the same total volume (the volume $Vol$ of ours is $0.0233$ and  raw input is $0.0237$). They are both composed of $8,000$ Voronoi cells with the same uniform beam thickness. Besides that, all external conditions for simulation are kept the same. The noisy tensor field leads to unstable microstructures with several artifacts as well as a poor static stress simulation. However, our smooth odeco tensor field produces much more stable microstructures and better static stress simulation result (i.e., with a smaller maximum Von Mises stress $S_{max}=1.47e6$ than the baseline $S_{max}=1.89e6$ and a smaller $P=6$ norm of Von Mises stress $S_{P_{norm}}=2.38e5$ than the baseline $S_{P_{norm}}=2.86e5$).  
\begin{figure}[h]
    \centering
    \includegraphics[width=\linewidth]{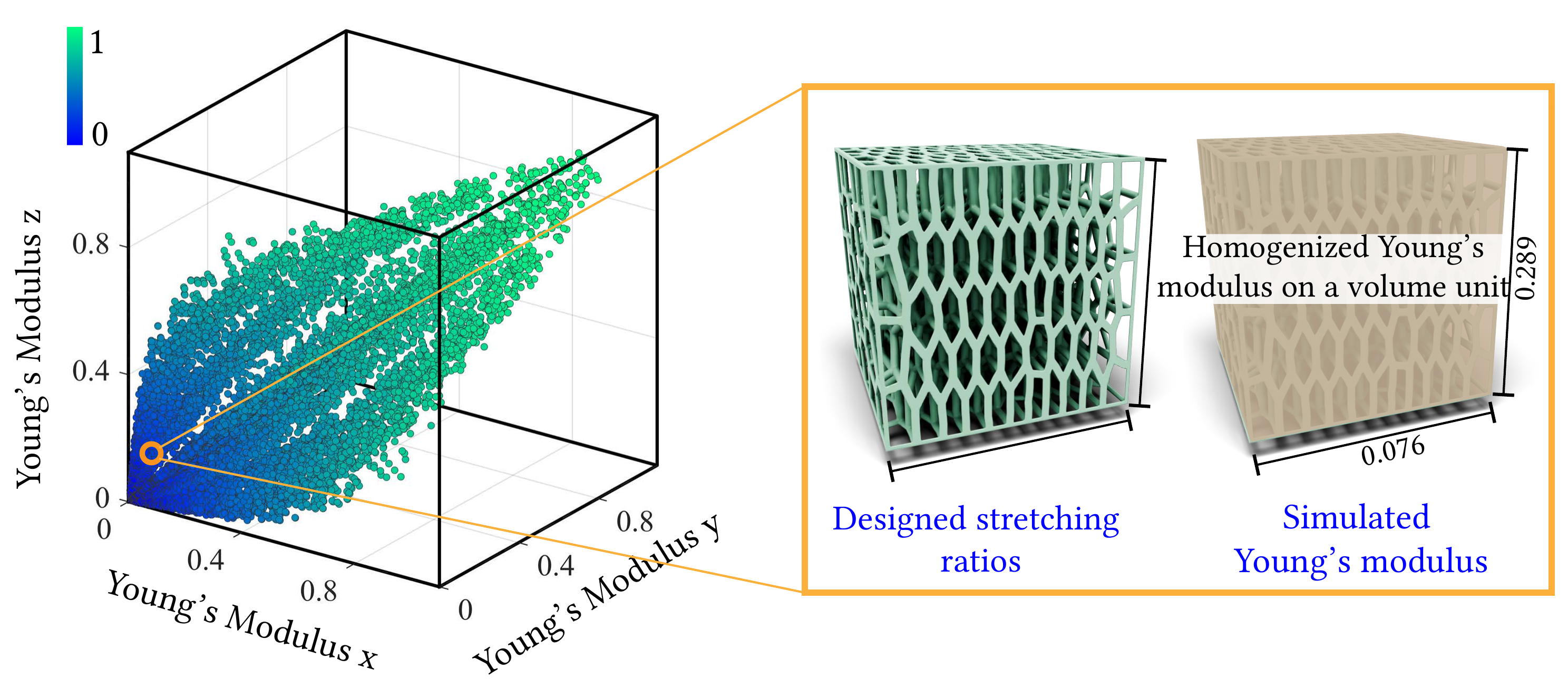}
  \caption{The material space that builds a relationship between the anisotropy of our microstructures and corresponding numerically homogenized Young's moduli.}
    \label{fig:fab_material}
\end{figure}

\begin{figure*}[h]
    \centering
    \includegraphics[width=\linewidth]{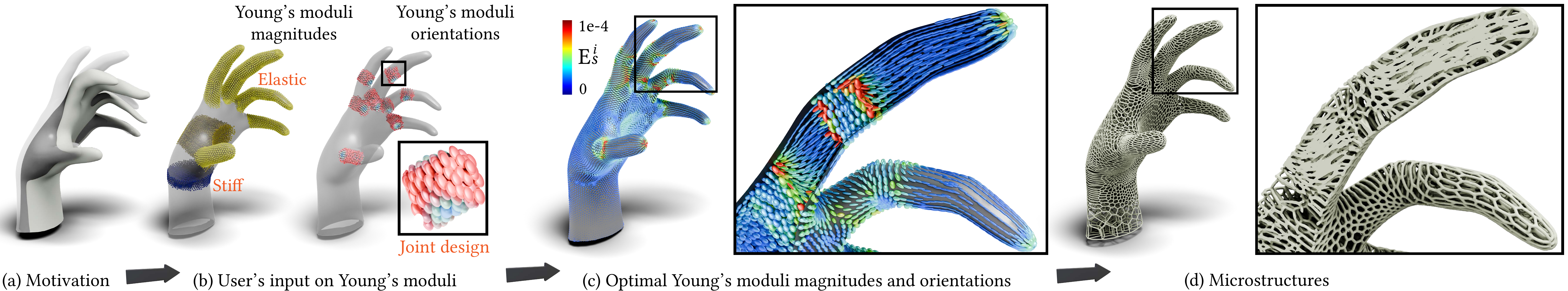}
    \caption{Fabrication of 3D open-cell microstructures for Hand model generated by user-designed anisotropic Young's moduli (elasticity). (a) A flexible and elastic hand is desired from a user; (b) Young's moduli input, where user-specified orientations and magnitudes are considered as the design guidance; (c) Optimal anisotropic Young's moduli with automatic shape conformity provided by our odeco tensor field method. The clipped views depict the interior of the tensor fields. 
  $E^i_{s}$ is the vertex-wise smoothness energy; (d) Generated microstructures which match the optimal Young's moduli. A retrieving procedure for anisotropic microstructure design in the material space is shown in Fig.~15.}
    \label{fig:fab_hand}
\end{figure*}

\subsection{Elastic Material Designs}
Elasticity designs are appealing for bioengineering fields, such as prosthetic artificial hand and bioprinted production. In this task, we only focus on the major properties of anisotropic elasticity (i.e., anisotropic Young's moduli). That is usually a popular and key component to design elastic structures in the literature. \revise{User-designed Young's moduli tensors typically exhibit different magnitudes in three orthogonal directions. Unlike rotational symmetry (RoSy) fields~\cite{vaxman2016directional}, our AOTF is a more reasonable representation for anisotropic Young's moduli.} 

We follow the idea of~\cite{fabrication2017martinez} to build a material space, which establishes a quantitative relationship between the parameters of microstructures and simulated Young's moduli, as shown in Fig.~\ref{fig:fab_material}. Then, we re-sample our material space for designing our own microstructures. The numerical simulated Young's moduli is implemented by the Fibergen library~\cite{fibergen2019}, which is a stable and fast Fourier transform-based homogenization method. The Young's moduli are simulated on representative volume units. In this material space, we are able to create the customizable microstructures which align with the user-designed anisotropic Young's moduli. Once the designed Young's moduli is given, we can retrieve the parameter settings of microstructure from that material space.

In Fig.~\ref{fig:fab_hand}, we demonstrate the procedure to design the anisotropic Young's moduli by our method. The anisotropic Young's moduli can be considered as AOTFs since they involve three orthogonal directions (locally) and corresponding Young's moduli. 
Motivated by the demand for artificial soft hands in robotics, our method streamlines the design process by requiring the magnitudes of Young's moduli as input. By default, the orientations of the optimized Young's moduli align with the boundary shape, simplifying design operations in most cases. Additionally, we introduce orientation constraints used for joint designs (e.g., to bend the fingers). The optimal Young's moduli automatically balance the shape conformity (especially for fingers) and user-specified constraints (for joints). After obtaining the optimal Young's modulus field, the tensors are normalized and projected onto the material space depicted in Fig.~\ref{fig:fab_material}. The final material is then modeled and fabricated based on the retrieved microstructure parameters. The generated microstructures nicely satisfy user's design requirement as shown in Fig.~\ref{fig:fab_hand}.

\section{Conclusion}
In this paper, we develop a new computational paradigm, AOTF, for generating complicated anisotropic tensor fields in volume. To our knowledge, this is the first odeco tensor method capable of designing smooth anisotropic tensor fields with several nice properties -- naturally aligning with the domain boundary and geometric features, as well as providing users with the flexibility of anisotropy control where needed. Our method demonstrates promising potential in terms of anisotropic meshing, microstructure fabrication, and material design.

\section{Limitations and Future Work}\label{se:future_work}
The proposed AOTF computation is a highly nonconvex problem, which could be difficult to solve because it is possible to get stuck in a local minimum. In this work, we propose a warm start-based joint optimization of tensor orientations and stretching ratios, which leads to a much faster convergence to a better local minimum, but it cannot guarantee to reach a theoretical global minimal solution. From our observations, while our AOTF results usually contain singularities with higher smoothness energy, during the optimization we mitigate these high energy regions associated with the singularities by optimizing both orientations and stretching ratios. This can relieve singularity issues. However, we have not considered the topology and structure of singularities in our tensor field design framework, which is beyond the scope of this work. In the future, we will improve the computational speed by using GPU-based or deep learning-based algorithm and implementation. 

Since 3D tensor-induced designs are essential for enhancing accuracy, efficiency, and stability in simulations across various fields, we will apply our method in quad or hex meshing, deformation, and texturing, etc. \revise{Especially, 3D anisotropic frame fields play a critical role in anisotropic hex meshing. However, to the best of our knowledge, there exists no tool that can generate anisotropic hex meshes from a tensor field, especially with large stretching ratios. The primary difficulty lies in ensuring that the odeco tensor field produces a singular graph that is meshable or integrable. Promisingly, the 2D planar results from \cite{couplet2024integrable} indicate that anisotropy significantly improves integrability, suggesting that a robust method for hex meshing could become feasible when extended to anisotropic cases. Therefore, our work on anisotropic frame field design and optimization strategies could serve as a good start to inspire future work on anisotropic hex meshing.}

\begin{acks}
The authors would like to thank the anonymous reviewers for their valuable comments and suggestions. Haikuan Zhu, Hongbo Li, Jing Hua, and Zichun Zhong were partially supported by National Science Foundation (OAC-1845962, OAC-1910469, and OAC-2311245), National Institutes of Health (R61NS119434), and General Motors research grant. 
\end{acks}

\bibliographystyle{ACM-Reference-Format}
\bibliography{ref}

\end{sloppypar}
\end{document}